\documentstyle[preprint,aps,prc,epsf,rotate]{revtex}
\tightenlines
\begin{document}
\input{epsf}
\def\Im{\mbox{\sl Im\ }}
\def\pd{\partial}
\def\oln{\overline}
\def\olft{\overleftarrow}
\def\ds{\displaystyle}
\def\bgreek#1{\mbox{\boldmath $#1$ \unboldmath}}
\def\sla#1{\slash \hspace{-2.5mm} #1}
\newcommand{\bra}{\langle}
\newcommand{\ket}{\rangle}
\newcommand{\vep}{\varepsilon}
\newcommand{\OBE}{{\mbox{\tiny OBE}}}
\newcommand{\met}{{\mbox{\scriptsize met}}}
\newcommand{\lab}{{\mbox{\scriptsize lab}}}
\newcommand{\cm}{{\mbox{\scriptsize cm}}}
\newcommand{\mcal}{\mathcal}
\newcommand{\Del}{$\Delta$}
\newcommand{\g}{{\rm g}}
\long\def\Omit#1{}
\long\def\omit#1{\small #1}
\def\beq{\begin{equation}}
\def\eeq{\end{equation} }
\def\bea{\begin{eqnarray}}
\def\eea{\end{eqnarray}}
\def\eqref#1{Eq.~(\ref{eq:#1})}
\def\eqlab#1{\label{eq:#1}}
\def\figref#1{Fig.~(\ref{fig:#1})}
\def\figlab#1{\label{fig:#1}}
\def\tabref#1{Table \ref{tab:#1}}
\def\tablab#1{\label{tab:#1}}
\def\secref#1{Section~\ref{sec:#1}}
\def\seclab#1{\label{sec:#1}}
\def\VYP#1#2#3{{\bf #1}, #3 (#2)}  % Volume, page (Year)
\def\NP#1#2#3{Nucl.~Phys.~\VYP{#1}{#2}{#3}}
\def\NPA#1#2#3{Nucl.~Phys.~A~\VYP{#1}{#2}{#3}}
\def\NPB#1#2#3{Nucl.~Phys.~B~\VYP{#1}{#2}{#3}}
\def\PRP#1#2#3{Phys.~Rep.~\VYP{#1}{#2}{#3}}
\def\ANP#1#2#3{Adv.~Nucl.~Phys.~\VYP{#1}{#2}{#3}}
\def\PL#1#2#3{Phys.~Lett.~\VYP{#1}{#2}{#3}}
\def\PLB#1#2#3{Phys.~Lett.~B~\VYP{#1}{#2}{#3}}
\def\PR#1#2#3{Phys.~Rev.~\VYP{#1}{#2}{#3}}
\def\PRC#1#2#3{Phys.~Rev.~C~\VYP{#1}{#2}{#3}}
\def\PRD#1#2#3{Phys.~Rev.~D~\VYP{#1}{#2}{#3}}
\def\PRL#1#2#3{Phys.~Rev.~Lett.~\VYP{#1}{#2}{#3}}
\def\FBS#1#2#3{Few-Body~Sys.~\VYP{#1}{#2}{#3}}
\def\AP#1#2#3{Ann.~Phys.~(N.Y.)~\VYP{#1}{#2}{#3}}
\def\ZP#1#2#3{Z.\ Phys.\  \VYP{#1}{#2}{#3}}
\def\ZPC#1#2#3{Z.\ Phys.\ C  \VYP{#1}{#2}{#3}}
\def\half{\mbox{\small{$\frac{1}{2}$}}}
\def\quarter{\mbox{\small{$\frac{1}{4}$}}}

% put your own definitions here:
\def\bge{\begin{equation}}
\def\ene{\end{equation}}
\def\bgea{\begin{eqnarray}}
\def\enea{\end{eqnarray}}
\def\nn{\nonumber}

%\oddsidemargin +0.0mm
%\evensidemargin +0.0mm
%\topmargin -13.0mm
%\pagestyle{myheadings}
%\voffset -1.5cm  % print at top of page
%\textheight 24cm
%\textwidth 17cm
%\oddsidemargin .5cm
%\evensidemargin .5cm

%%%%%%%%       {\bf DRAFT! }
%%%%%%%%        \preprint{HEP/123-qed}

\title{Coherent Compton scattering on light nuclei 
in the $\Delta$-resonance region}

\author{L.\ Van Daele$^a$, A.\ Yu.\ Korchin$^{a,b}$\footnote{Permanent address:
National Science
Center `Kharkov Institute of Physics and Technology', 61108 Kharkov,
Ukraine}, D.\ Van Neck$^a$, O.\ Scholten$^b$, and M.\ Waroquier$^a$}
%%%%%% \homepage{http://www.Second.institution.edu/~Charlie.Author}

\address{$^a$ Department of Subatomic and Radiation Physics, University of
Gent, B-9000 Gent, Belgium}
%%%%% \email{Second.Author@institution.edu}

\address{$^b$ Kernfysisch Versneller Instituut, University of Groningen, 9747 AA
Groningen, The~Netherlands}

%%%%%%   \date{\today}   % It is always \today, today, but you may specify any date with \date.

\maketitle

\begin{abstract}
Coherent Compton scattering on light  nuclei in the
$\Delta$ resonance region is studied in the impulse approximation and
is shown to be a sensitive probe of the in-medium properties of 
the \Del\ resonance. The elementary amplitude on a single nucleon 
is calculated from the unitary K-matrix approach developed
previously. Modifications of the properties of the $\Delta$ resonance 
due to the nuclear medium are accounted for through the self-energy 
operator of the $\Delta$, calculated from the one-pion loop. 
The dominant medium effects such as the Pauli blocking, mean-field 
modification of the nucleon and $\Delta$ masses,  and particle-hole
excitations in the pion propagator are consistently included in
nuclear matter.

\end{abstract}
{\bf 2001 PACS} numbers:  25.20.Dc, 13.60.Fz\\
%%%%%% \pacs{}    %PACS, the Physics and Astronomy Classification Scheme.

\maketitle

%%%%%%%       \tableofcontents

\section{Introduction}

Coherent Compton scattering on nuclei in the region of the $\Delta$ resonance
is of considerable interest.  The reaction allows one to study the
propagation and decay
of the $\Delta$ in the nuclear medium. In particular the shift of the
pole position and a  change of the
width of the $\Delta$ peak, reflect sensitively
 in the cross section and polarization observables.

 For a comprehensive review on Compton scattering we refer to the recent
reference~\cite{Hut00}. Here we only
mention several models relevant for the present study. The so-called
''schematic
and extended'' models~\cite{Hay84,Kra98} made use of the optical
theorem and dispersion relations to relate the Compton-scattering amplitude with the total
photoabsorption cross section.
Important contributions to the amplitude, such as nuclear kinetic and meson
seagull
terms, and $\Delta$ excitation, were added explicitly~\cite{Kra98}.
The data for Compton scattering were described quite well at forward scattering angles.

Compton scattering was extensively studied in the $\Delta$-hole
model~\cite{Kra98,Kor94,Pas96} which was originally developed by Koch, Moniz
and Ohtsuka in~\cite{Koc83,Koc84}. The $\Delta$ excitation is included through
the
$M1$ transition, the background is represented by the Kroll-Ruderman part
of the virtual-pion photoproduction amplitude, while in Refs.~\cite{Kra98,Pas96}
the proton Thomson term (with $E1$ multipolarity) is also added.

The starting point of the present model is  the full Compton-scattering
amplitude $T_N$ on a free nucleon
that includes all dominant mechanisms at energies up to, at least, the $\Delta$
resonance.
This amplitude has been  obtained in the unitary K-matrix approach, similar to the
calculations presented in refs.~\cite{Sch96,Kor98} for pion-nucleon scattering,
pion photoproduction and Compton scattering on the proton. Parameters of
the model are fixed from a detailed fit to observables for the above
elementary reactions. The Thomson limit dictated by the low-energy theorem on the
nucleon and the
Kroll-Ruderman term of the photoproduction amplitude are automatically included.

Medium effects are taken into account via the self-energy of 
the $\Delta$,
 while the self-energy is calculated from the pion-loop
mechanism embedded in nuclear matter,
as described in detail in sect.II. 
In particular, Pauli blocking effects, mean-field modifications
of the nucleon and $\Delta$ masses,
and particle-hole contributions to the
pion propagator are consistently calculated.
Thus we avoid arbitrariness associated  with introducing a
 parameterized self-energy.

 The amplitude on a nucleus is calculated in
 the impulse approximation.  A formfactor is
introduced to account for finite size effects.
The specific density dependence of the medium effects
 is accounted for through a local density approximation.
Fermi-motion is treated in the so-called
factorization approximation, 
in which the elementary Compton-scattering amplitude is
calculated on a nucleon with an effective momentum
\cite{Tia80,Era90,Dre99b,Lan73}.
The latter is taken in such a way that
the energy-momentum conservation for the $\gamma \, N$ 
scattering holds, thus
ensuring gauge invariance at the one-body level.

Differential cross sections and photon asymmetries are calculated
for light nuclei ($^4$He and $^{12}$C) in the 
energy region 150 -- 250 MeV.

The structure of the paper is as follows. In sect.II we describe
the calculation of medium effects in the $\Delta$ propagator.  Sect.III
presents the basic formalism for calculating 
the Compton-scattering amplitude
on nuclei in the impulse approximation.
Results for the cross section and photon asymmetry 
are presented in sect.IV.
Conclusions and prospects are further outlined.
In Appendix A formulas for the pion self-energy are collected.
Details of the kinematics involved in the reaction on a nucleus
are contained in Appendix B.

\section{Medium modifications of the \Del\ resonance}
\label{sec: medium}

The properties of the $\Delta$ in the nuclear medium
are calculated in a relativistic framework for symmetrical
nuclear matter, along the lines of Refs.~\cite{Herb1,Kim1,Liu1}.
The medium modifications, which are expressed through the dressing of
the $\Delta$ propagator, are investigated using different levels 
of approximation.
The imaginary part of the $\Delta$ self-energy
(or the $\Delta$ decay width) is calculated in different models
for the nuclear medium. Dispersion relations are used to determine the
real part (mass modification) of the $\Delta$ self-energy.

In a first subsection the Rarita-Schwinger formalism is introduced
to describe the $\Delta$ resonance. The next three subsections
deal with the $\Delta$ decay width. In a first step
the nuclear medium is presented as a non-interacting 
Fermi-sea of nucleons
leading to the calculation of the Pauli-corrected decay width and the
spreading width. Subsequently mean-field effects are included within the
$\sigma\omega-$model of Walecka \cite{Wal74}.
Throughout, the conventions of Bjorken and Drell \cite{Bjor} 
are used, and all expressions are derived in the nuclear matter
rest frame.

\subsection{Formalism for the $\Delta$ self-energy}
\label{subsec:self-energy}

The free $\Delta$ propagator in the Rarita-Schwinger
formalism has the following form \cite{Rari}:
\bgea
G_{\Delta}^0 (p_{\Delta})^{\mu\nu}  &=&
\frac{1}{ \rlap{/}p_{\Delta}
- M_{\Delta} +i\epsilon}\,
\Biggl[\, g^{\mu\nu} - \frac{1}{3}\gamma^{\mu}\gamma^{\nu}
- \frac{2p_{\Delta}^{\mu} p_{\Delta}^{\nu}}{3M_{\Delta}^{2}}
-\frac{\left(\gamma^{\mu} p_{\Delta}^{\nu} -\gamma^{\nu}
p_{\Delta}^{\mu}\right)}{3M_{\Delta}} \, \Biggr]
\enea
with the mass $M_{\Delta} = 1232$ MeV
and $p_\Delta = ( p_\Delta^0 ,\, \vec{p}_\Delta)$.  The 
invariant mass of the $\Delta$ resonance will be denoted 
by $W_\Delta = (p_\Delta^2)^{1/2}$.

The propagator has to be dressed due to the
interaction of the $\Delta$ with the nuclear medium. This
is taken into account by introducing the self-energy in the equation
for the inverse propagator
\bea
 ( G_\Delta^{-1} )^{\mu\nu}
&=&( ( G^0_\Delta )^{-1} )^{\mu\nu}
 - \Sigma_\Delta^{\mu\nu} =
( \rlap{/}p_{\Delta} -M_\Delta )\,
( {\cal P}^{3/2} )^{\mu\nu} \nonumber\\
 &-& 2( \rlap{/}p_{\Delta} - M_\Delta )\,
( {\cal P}^{1/2}_{11} )^{\mu\nu}
+\sqrt{3} M_\Delta [\, ( {\cal{P}}^{1/2}_{12} )^{\mu\nu} +
( {\cal P}^{1/2}_{21} )^{\mu\nu}\, ]
- \Sigma_{\Delta }^{\mu\nu} \,,
\eea
where the inverse free propagator is written in terms of the
projection operators on the spin-3/2 and spin-1/2
sectors~\cite{Benm1}
($\, {\cal P}^{3/2},\, {\cal P}^{1/2}_{ij},\,
ij=11,\, 22,\, 12,\, 21\,$).
In this paper we concentrate on the dominant 
spin-3/2 component of the $\Delta $
propagator, and therefore the self-energy is taken as
\bge
\Sigma_{\Delta }^{\mu\nu} = \Sigma_{\Delta }^{3/2}
 \;({\cal{P}}^{3/2})^{\mu\nu},
\ene
where we discarded terms proportional to the spin-1/2 
projection operators.
The spin-3/2 component of the self-energy has the structure
\beq
\Sigma_{\Delta}^{3/2} =
C_{\Delta} ( p_{\Delta} ) + D_{\Delta} (p_{\Delta} )
\,\rlap{/}{p_{\Delta}}   \,,
\eqlab{sigma-3/2}
\label{sigma-3/2}
\eeq
and the dressed propagator becomes
\bea
G_{\Delta} ( p_{\Delta} )^{\mu\nu} &=&
\frac{1}{ \rlap{/}p_{\Delta} - M_{\Delta} -
\Sigma_{\Delta}^{3/2}  }
\ ( {\cal P}^{3/2} )^{\mu\nu}
\nonumber\\
 &=&
[  {\rlap{/}p_{\Delta}} ( 1 - Re D_{\Delta} (p_{\Delta} ) )
- ( M_{\Delta} + Re C_{\Delta} (p_{\Delta} ) )
\nn\\
&-& i ( Im C_{\Delta} (p_{\Delta} ) +
Im D_{\Delta}(p_{\Delta} ) \rlap{/}{p_{\Delta}} ) ]^{-1}
\  ( {\cal{P}}^{3/2} )^{\mu\nu}.
\label{med-mod-de-prop}
\eea
Note that in vacuum $C_{\Delta} ( p_{\Delta} )$ and $D_{\Delta}
 ( p_{\Delta} )$ are functions of the invariant $W_\Delta$ only, 
while in the medium these functions acquire additional 
dependencies on the 3-momentum
$|\vec{p}_\Delta|$ and the nuclear density $\rho$.
We keep $p_\Delta$  as an argument in
Eq.~(\ref{sigma-3/2}) for the general situation.

\subsection{Width of the $\Delta$ resonance and Pauli blocking}
\label{subsec:Pauli}

The imaginary part of the $\Delta$ self-energy is calculated 
from an effective interaction Lagrangian.
We focus on the pion-nucleon decay channel in a nuclear environment.
The $\pi N\Delta$ interaction Lagrangian is taken in the form
\cite{Benm1,Nath1,Olss1}
\bge
{\cal L}_{\pi N\Delta} = \frac{f_{\pi
N\Delta}}{m_{\pi}} \ g^{\mu\nu} \overline{\Delta}_{\mu}
\vec{T}^\dagger  \Psi \
(\partial_{\nu} \vec{\pi} )\ +\ H.c. \,,
\ene
where $\Psi$ ($\Delta_\mu$) is the nucleon ($\Delta$) field operator,
$\vec{\pi}$ the pion field, and ${T}_i$ is the
isospin $\frac{1}{2} \leftrightarrow \frac{3}{2}$
transition operator with the normalization condition
$
{T}_i {T}^{\dagger}_j = \delta_{ij} - \frac{1}{3}{\tau}_i
{\tau}_j\,, \;\;\; ( i,j = 1,2,3)\,.
$
In the above Lagrangian we dropped the off-shell
coupling of the $\Delta$ described by the parameter $z_\pi$
\cite{Benm1,Nath1,Olss1}, since
it affects only the spin-1/2 components which are not
considered in the present work.

The width of the resonance is defined by the imaginary part of the
self-energy as
\bge
\Gamma_{\Delta}(p_{\Delta})= -2 Im  \,
[ \, C_{\Delta}(p_{\Delta}) + W_\Delta D_{\Delta}(p_{\Delta}) \,].
\label{general_width_1}
\ene
This can more formally be written as a trace
\bgea
\Gamma_{\Delta}(p_{\Delta})= -2 Im  \, \left[
\frac{1}{4}\, Tr \, \left(\, \Sigma_{\Delta}^{\mu\nu}
\,({\cal P}^{3/2})_{\nu\mu}\,
\frac{ (\rlap{/} {p}_\Delta + W_\Delta)}{2 W_\Delta}\,\right)
\right]
\label{S3/2}
\enea
The contribution to the self-energy for a $\Delta$ decaying into
a nucleon and a pion is
\bge
-i\left(\Sigma_{\Delta}\right)_{\mu\nu}=\Big( \frac{f_{\pi N\Delta}}{m_\pi} \Big)^2
 \sum_{a=1,2,3}
\int\frac{d^{4}k}{(2\pi)^4}
q_{\mu} T^\dagger_a
G_N(k) (- q_{\nu}) T_a D_{\pi}(q),
\label{general self-en}
\ene
where $q=p_\Delta - k$, and $G_N(k)$ and $D_{\pi}(q)$ are
the nucleon and pion propagators, the structure of which
depends on the model used for the nuclear medium.
We will now discuss different approximations and their implications
for the $\Delta$ decay width.

In vacuum the nucleon and pion propagators in
(\ref{general self-en}) reduce to the free nucleon propagator
\bge
G_{N}^{0}(k)=\frac{1}{ \rlap{/} {k}- M_{N} + i\epsilon}=
\frac{M_{N}}{E_{\vec{k}}}
\left(\frac{\Lambda^{+}(\vec{k})}{k_{0}-E_{\vec{k}}+i\epsilon}
- \frac{\Lambda^{-}(-\vec{k})}{k_{0}+E_{\vec{k}}-i\epsilon} \right)
\label{nucl-prop-vrij}
\ene
and the free pion propagator
\bge
D_{\pi}^0 (q) = \frac{1}{q^2 -m_{\pi}^2 +i\epsilon} \,,
\ene
where $E_{\vec{k}} = \sqrt{|\vec{k}|^{2} + M_{N}^{2}}$ and
$\Lambda^{\pm}$ are the projection operators
\bgea
\Lambda^{\pm}( \pm \vec{k}) = \frac{ \pm \gamma_{0}
E_{\vec{k}}-\vec{\gamma}\cdot\vec{k}\ +
M_N }{ 2 M_N }\,.
\eqlab{proj3}
\label{proj3}
\enea

Applying the Cutcosky rules for the discontinuity of the loop
integral in Eq.(\ref{general self-en}), and using
$\, \sum_{a=1,2,3}T^{\dagger}_a T_a = 1\,$
one gets the conventional expression for the width in vacuum
\bge
\Gamma_{\Delta}^{vac} = \frac{f_{\pi N\Delta}^{2}
\left(E_{{k}_{\pi}} + M_N \right)}{12\pi m_{\pi}^2 W_{\Delta}}
{k}_{\pi}^3\,,
\label{vac-width}
\ene
where
\bge
{k}_{\pi}^{2} = \frac{1}{4 W_\Delta^2} \, \label{k2}
{[ W_\Delta^2 - (M_N +m_\pi)^2 ] [ W_\Delta^2 - (M_N -m_\pi)^2 ] }
\ene
and $E_{{k}_{\pi}} = \sqrt{{k}_{\pi}^{2} + M_{N}^{2}}$.
Putting $W_\Delta = M_\Delta$ in  Eq.~(\ref{vac-width}) and using
the experimental value ${\Gamma}^{vac}_{\Delta} = 115$ MeV
one obtains the coupling constant $f_{\pi N{\Delta}} = 2.15$.

The effect of Pauli blocking is included by replacing
the free nucleon propagator by a medium-modified one.
If one models the nuclear environment as a non-interacting
Fermi-sea of nucleons the propagator becomes
(see, e.g.,~\cite{Serot1})
\bgea
G_{N}^{FG} (k) &=& \left[ \frac{\theta(|\vec{k}|-k_{F})}
{\rlap{/} {k}- M_{N} + i\epsilon} +
\frac{\theta(k_F-|\vec{k}|) }{ \rlap{/} {k}- M_{N}-i\epsilon}
\right] \theta(k^0) +
\frac{\theta(-k^0)}{\rlap{/} {k}- M_{N} + i\epsilon}
\nonumber\\
&=& \frac{M_{N}}{E_{\vec{k}}}
\left(\Lambda^{+}(\vec{k}) \left(
\frac{\theta(|\vec{k}|-k_{F})}{k_{0}-E_{\vec{k}} +i\epsilon}
+\frac{\theta(k_{F}-|\vec{k}|)}{k_{0}-E_{\vec{k}} -i\epsilon} \right)
- \frac{\Lambda^{-}(-\vec{k})}{k_{0}+E_{\vec{k}}-i\epsilon} \right)\,.
\eqlab{Fermi-gas-prop}
\enea
This is written in the nuclear-matter rest frame, an 
explicitly covariant form is discussed in~\cite{Herb1}.
Implementing this propagator in Eq.~(\ref{general self-en}) and 
defining the Fermi-energy $\,E_F = \sqrt{k_F^2 + M_N^2}$
and $\,E_{\pm} = (p_\Delta^0 E_{k_\pi} \pm 
|\vec{p}_\Delta | k_\pi ) / W_\Delta$
we obtain the Pauli-corrected $\Delta $ decay width
\begin{eqnarray}
\Gamma_{\Delta}^{D}(p_{\Delta})= \frac{f_{\pi N\Delta}^{2}
\left(E_{{k}_{\pi}} + M_N \right){k}_{\pi}^2 }{24\pi m_{\pi}^2
|\vec{p}_{\Delta}|}  \, [E_+ - \mbox{max}(E_- ,E_F )]\ \theta(E_+ -E_F )\,.
\label{PC-width}
\end{eqnarray}
This result can be separated into three different energy regions:
%%%%%%
\begin{enumerate}
\item $E_- > E_F$ : the Pauli-corrected decay width
reduces to the vacuum expression (\ref{vac-width}), making the width
$|\vec{p}_{\Delta}|$-independent;
\item  $E_+ > E_F > E_-$ :
\bge
\Gamma_{\Delta}^{D}(p_{\Delta})= \frac{f_{\pi N\Delta}^{2}
\left(E_{{k}_{\pi}} + M_N \right)}{24\pi m_{\pi}^2
W_{\Delta}|\vec{p}_{\Delta}|} {k}_{\pi}^2 ( E_{{k}_{\pi}}p_{\Delta}^0
+|\vec{p}_{\Delta}|{k}_{\pi} - W_{\Delta} E_F ) \,.
\ene
The width becomes $|\vec{p}_{\Delta}|$- and density-dependent.
Taking the limit $|\vec{p}_{\Delta}| \rightarrow 0$ reduces
the width to its vacuum value if $E_{k_{\pi}} > E_F$,
and to zero if $E_{k_{\pi}} \leq E_F$;
\item $E_F > E_+$ : the width becomes zero
($\Gamma_{\Delta}^{D}(p_{\Delta}) = 0$).  This means the
Fermi-sphere engulfs the decay sphere completely, making the
Pauli blocking complete.
\end{enumerate}
%%%%%

In \figref{ds_pc_vm}(a) the full dependence on $\rho$ 
and $|\vec{p}_{\Delta}|$ is shown for $W_{\Delta} = 1232$ MeV. 
The results at densities 1.2, 0.8, 0.4 and 0.05 times normal nuclear 
matter density $\rho_{0}$ ($k_F^0 = 1.333 fm^{-1}$) are plotted
separately in \figref{ds_pc_vm}(b).
At high $\Delta$-momenta and low densities the energy of the decay nucleon  
lies well above the Fermi energy, and no blocking occurs. At somewhat 
lower momenta of the $\Delta$ part of the momenta of the decay 
nucleon are Pauli blocked. 
With increasing density this blocking may become complete
for the lowest $\Delta$-momenta making the $\Delta$
unable to decay into a pion-nucleon pair (see dashed and 
dotted curves in \figref{ds_pc_vm}(b)). 
These phase-space considerations result in a 
strong energy-dependence of the $\Delta$ decay width.

%%%%%%%%%%%%%%%%%%
\subsection{Spreading width}
\label{Spreading}

In the nuclear medium the pion will strongly interact with
the surrounding baryons creating nucleon-hole
and $\Delta$-hole excitations. This can be taken into account
by dressing the pion propagator with the proper pion self-energy
\bge
D_{\pi} (q) = \frac{1}{q^2 - m_{\pi}^2- \Pi_{\pi}(q) + i\epsilon}\ ,
\label{dressed-pp}
\ene
where $\Pi_{\pi}(q) = \Pi_{ph}(q) + \Pi_{{\Delta}h}(q)$ is
the polarization self-energy of the pion.
In our calculations we use the pion-nucleon pseudo-vector coupling
\bge
{\cal L}_{{\pi}NN} = \frac{f_{{\pi}NN}}{m_{\pi}} \overline{\Psi}
{\gamma}^{\mu} {\gamma}_{5} \vec{\tau} {\Psi} \, (\partial_{\mu}
\vec{\pi})
\ene
with the $\pi NN$-coupling constant $f_{\pi NN} = 1.01$ \cite{Kim1}.
In this work we limit ourselves to forward and backward scattered
particle-hole excitations, and omit
anti-nucleon excitations in \eqref{Fermi-gas-prop}, which play a role
only for very large pion momenta. Also intermediate $\Delta$-hole states
are omitted from the pion self-energy. We expect that their contribution
is more suppressed than the estimate in ref.\cite{Herb1} 
when the width of the $\Delta$ resonance is taken into account.
In principle, a complete
calculation of the $\Delta$-hole states would
require self-consistency between the pion and $\Delta$ self-energies,
which falls outside the scope of the present paper.

The lowest-order pion self-energy now reads
\bgea
\Pi_{ph}^0 (q) &=& -i \left(\frac{f_{{\pi}NN}}{m_{\pi}}\right)^2
2 \int \frac{d^4k}{(2 \pi)^4} Tr \left[\rlap{/}q {\gamma}_{5}
(\hat{\rlap{/}l} + M_N) \rlap{/}q {\gamma}_{5} (\hat{\rlap{/}k}
+ M_N) \right]\nn\\
\mbox{} &&\ \ \times\ \frac{1}{4E_{\vec{k}+\vec{q}} E_{\vec{k}}}
\left[\frac{\theta(|\vec{k}+\vec{q}| - k_F)}{(k_0+q_0 -
E_{\vec{k} + \vec{q}} + i\epsilon)} \frac{\theta(k_F - |\vec{k}|)}
{(k_0 - E_{\vec{k}} - i\epsilon)} \right. \nn\\
\mbox{} && \left.\ \ \ \ \ \ \ \ \ \ \
+\ \ \ \frac{\theta(k_F - |\vec{k}+\vec{q}|)}
{(k_0 + q_0 - E_{\vec{k}+\vec{q}} - i\epsilon)}
\frac{\theta(|\vec{k}|- k_F)}
{(k_0 - E_{\vec{k}} + i\epsilon)} \right]
\label{lowest-order-phps}
\\
&=& 8 \left( \frac{f_{{\pi}NN}}{m_{\pi}}\right)^2
\left[|\vec{q}|^2 C_0(|\vec{q}|) - C_2(|\vec{q}|)
+ 2M_N^2 q^2 L_0(q) \right]\,,
\label{low-order-phps}
\enea
where $\hat{k} = (E_{\vec{k}}, \vec{k})$ and $\hat{l} =
(E_{\vec{k} + \vec{q}}, \vec{k}+\vec{q})$. The analytical expressions 
for the functions $L_0$, $C_0$ and $C_2$ are given in Appendix~A.
When summing the series of particle-hole bubbles the 
effects of short-range correlations are important.
These short-range correlations are accounted for  in the standard 
way by introducing the Landau-Migdal parameter $g_{NN}'
= 0.6$ \cite{Suzuki1},
\bge
\Pi_{ph}(q) = q^2 \frac{\Pi_{ph}^0 (q)}{q^2 + g_{NN}'
\Pi_{ph}^0 (q)}    \,.
\label{src-phps}
\ene
Using the pion self-energy ${\Pi}_{ph}$ in the pion
propagator from (\ref{dressed-pp}) we get the
spreading width of the $\Delta$ resonance in the medium
\bgea
\Gamma_{\Delta}^{S} (p_{\Delta}) = \left(\frac{2}
{3W_{\Delta}^3}\right)\left(\frac{f_{{\pi}N{\Delta}}}{m_{\pi}}
\right)^2 \int_{k_F}^{\sqrt{(p_{\Delta}^0)^2 - M_N^2}}
 d|\vec{k}| \frac{|\vec{k}|^2}
{(2\pi)^2} \int_{-1}^{1} d \cos\theta \left(\hat{k}\cdot p_{\Delta}
+ M_N W_{\Delta}\right)\nn\\ \times
\left(W_{\Delta}^2 \hat{q}^2 - (p_{\Delta}\cdot \hat{q})^2 \right)
\frac{1}{2E_{\vec{k}}} \; \frac{Im \Pi_{ph}(\hat{q})\;  
\theta(-\hat{q}^2)} {[\hat{q}^2 - m_{\pi}^2
- Re \Pi_{ph}(\hat{q})]^2 + [Im \Pi_{ph}(\hat{q})]^2}
\eqlab{spread}
\label{spreading-width}
\enea
where $\hat{q} = p_{\Delta} - \hat{k}$.
Since the pion self-energy
only receives contributions from the particle-hole excitations, 
its imaginary part is non-zero only for 
space-like pion momenta. This puts
restrictions on the integration boundaries in \eqref{spread}.
The two-dimensional integral was evaluated numerically.
As an example we have plotted in the upper panel of
\figref{ps_im} the imaginary part of the
pion self-energy in function of the pion momentum 
$|\vec{q}|$, as it appears in \eqref{spread}
for the specific $\Delta$ kinematics : $W_{\Delta} = 1232$ MeV,
$|\vec{p}_{\Delta}| = 200$ MeV and pion energy 
$q^0 = (p_{\Delta}^0 - E_F)/2$.

The results for the spreading width are shown 
in \figref{ds_phsrc_vm}.
The spreading width is roughly proportional to the density, 
which can be understood on the basis of the phase space 
available for the hole states.
As can be seen from \figref{ds_phsrc_vm}(b) it is only weakly 
dependent on the 3-momentum $\mid\vec{p}_{\Delta}\mid$. 
Also the dependence on $W_{\Delta}$ turns out to be rather weak
in the region of interest.
The total width of the $\Delta$ in this non-interacting
Fermi-sea of nucleons is given by the sum of this spreading width and
the Pauli-corrected decay width from the previous section.

%%%%%%%%%%%%%%%%%%
\subsection{Mean-field effects in the nucleon and $\Delta$ 
self-energy }

\label{Effective mass}

A refinement to the free Fermi-gas model can be made using
the Walecka $\sigma\omega-$model~\cite{Wal74}
in the mean-field approximation.
Here the $\sigma-$ and $\omega-$meson couple to the nucleon
resulting in the mean scalar and vector fields
$\langle \Phi_s\rangle $ and $\langle V^{\mu}\rangle $.

In the nuclear matter rest frame the spatial part of 
$\langle V^{\mu}\rangle $ is averaged to zero, and the constant
mean-field contribution to the nucleon self-energy becomes
\bgea
\Sigma_{N} (k) = -g_s^N \Phi + g_v^N V {\gamma}_0 ,
\enea
where $g_s^N$ and $g_v^N$ are the coupling constants of the 
scalar and vector field with masses $m_s$ and $m_v$
respectively, and $\langle \Phi_s\rangle = \Phi $,
$\langle V^{\mu}\rangle = {\delta}_{\mu 0} V$.
This self-energy can be implemented in the modified Dirac 
equation~\cite{Wal74}
\bgea
&& \left[ \rlap{/}{k} - M_N  - \Sigma_N(k) \right]
\Psi_N (k)= 0 \,.
\label{med-nucl-prop-2}
\enea
Introducing the effective nucleon four-momentum $k^{\star} =
(k^0 - g_v^N V, \vec{k})$ and mass $M_N^{\star} = M_N - g_s^N \Phi$,
the nucleon spectrum is modified to ${k^{\star}}^2 = 
{M_N^{\star}}^2$, or $k_0 = g_v^N V \pm E_{\vec{k}}^{\star}$,
where $E_{\vec{k}}^{\star} =\sqrt{|\vec{k}|^2 + (M_N^{\star})^2}\,$.

In order to assess the sensitivity of the results to the mean-field
parameters we have performed calculations taking 2 parameter sets
from \cite{Serot1}, henceforth called set I and II. Set I, called
QHD-I in \cite{Serot1}, results from a pure mean-field 
approximation to the binding energy. The dimensionless 
ratios of coupling constants and meson masses have values
$C_s^2 = (g_s^N M_N / m_s)^2 = 267.1$, 
$C_v^2 = (g_v^N M_N / m_v)^2 = 195.9$. The nuclear matter
equilibrium density is at $k_F^0 = 1.42 fm^{-1}$, with
binding energy 15.75 MeV and an effective nucleon
mass $M_N^{\star} / M_N = 0.56$ at ${\rho}_0$.
The full density dependence of the effective mass is defined
explicitly by the self-consistency equation
\bge
M_N^{\star}= M_N - \frac{(g_s^N)^2}{m_s^2}\frac{M_N^{\star}}{\pi^2}
\left[ k_F E_F^{\star} - (M_N^{\star})^2 \ln
\left(\frac{k_F + E_F^{\star}}{M_N^{\star}}\right)
\right]\label{mneff-mn}.
\ene
where $E_{F}^{\star} =\sqrt{k_{F}^2 + (M_N^{\star})^2 }\,$.
Set II, called the relativistic Hartree approximation in \cite{Serot1},
takes into account vacuum fluctuation corrections
to the binding energy. The parameters are $C_s^2 = 228.2$, 
$C_v^2 = 147.8$. The equilibrium density is taken at
$k_F^0 = 1.30 fm^{-1}$, with a binding energy of 15.75 MeV
leading to an effective nucleon mass $M_N^{\star} / M_N = 0.73$
at equilibrium density and the self-consistency equation reads :
\bgea
M_N^{\star}= M_N && - \,\, 
\frac{(g_s^N)^2}{m_s^2}\frac{M_N^{\star}}{\pi^2}
\left[ k_F E_F^{\star} - (M_N^{\star})^2 \ln
\left(\frac{k_F + E_F^{\star}}{M_N^{\star}}\right)
\right] \nn\\
&+& \frac{(g_s^N)^2}{m_s^2}\frac{1}{{\pi}^2}
\left[ {M_N^{\star}}^3 \ln \left(\frac{M_N^{\star}}{M_N}\right)
- {M_N^2} (M_N^{\star} - M_N) 
\right. \nn\\
\mbox{ }&&
\left.
 - \,\,\, \frac{5}{2}{M_N}(M_N^{\star} - M_N)^2
- \frac{11}{6}(M_N^{\star} - M_N)^3
\right]
\label{mneff-mn-2}.
\enea

The full density dependence of 
the effective nucleon masses in both cases are shown in 
\figref{kf_efmn_1}. We see a strong reduction of the effective 
nucleon mass compared to the free nucleon mass
with increasing density in both cases, with
a slower decline 
when the influence of negative energy-states is considered.

In the extended mean-field model of ref.~\cite{Herb1}
the $\Delta$ is assumed to move in the mean $\sigma$ 
and $\omega$ fields. The mean-field contributions to the
$\Delta$ self-energy can be treated in an analogous way
as for the nucleon, i.e. they are absorbed in the effective 
$\Delta$ mass $M_{\Delta}^{\star}$ and 4-momentum 
$p_{\Delta}^{\star}$. In this paper we employ the 
so-called universal couplings \cite{Herb1}:
$g_s^N = g_s^{\Delta}$ and $g_v^N = g_v^{\Delta}$ and as a result
the $\Delta$ effective mass $M_{\Delta}^{\star}(\rho) $
may be expressed as $M_{\Delta}^{\star}(\rho) = M_{\Delta} - (M_N -
M_N^{\star})$.

We can now investigate the influence of these mean-field
modifications on the Pauli-corrected decay width and the spreading
width of the $\Delta$ by replacing the Fermi-gas nucleon propagator
(\eqref{Fermi-gas-prop}) in the expression (\ref{general self-en}) and 
in the calculation of the pion self-energy (see Eq.(\ref{lowest-order-phps}))
with the mean-field propagator
\bgea
G_{N}^{\sigma\omega} (k) &=& \left[ \frac{\theta(|\vec{k}|-k_{F})}
{\rlap{/} {k^{\star}}- M_{N}^{\star} + i\epsilon} +
\frac{\theta(k_F-|\vec{k}|) }{ \rlap{/} {k^{\star}}- M_{N}^{\star}
-i\epsilon}
\right] \theta(k^{\star}_0) +
\frac{\theta(-k^{\star}_0)}{\rlap{/} {k^{\star}} - M_{N}^{\star}
+ i\epsilon}
\nonumber\\
&=& \frac{M_{N}^{\star}}{E_{\vec{k}}^{\star}}
\left( {{\Lambda}^{{\star}+}}(\vec{k}) \left(
\frac{\theta(|\vec{k}|-k_{F})}{k_{0}^{\star} - E_{\vec{k}}^{\star}
+i\epsilon}
+ \frac{\theta(k_{F}-|\vec{k}|)}{k_{0}^{\star} - E_{\vec{k}}^{\star}
 -i\epsilon} \right)
- \frac{{\Lambda^{{\star}-}}(-\vec{k})}
{k_{0}^{\star} + E_{\vec{k}}^{\star} -i\epsilon} \right)\,,
\eqlab{s_o-Fermi-gas-prop}
\enea
where ${\Lambda}^{{\star}\pm}$ follows from~\eqref{proj3} with the
nucleon energy $E_{\vec{k}}$ and the nucleon mass $M_N$ 
replaced by the effective variables $E_{\vec{k}}^{\star}$ 
and $M_N^{\star}$. The resulting expressions are formally identical
to the free Fermi-gas expressions, if all kinematical
quantities are replaced by their effective equivalents.
In particular, we introduce the effective in-medium
mass of the $\Delta$, $W_{\Delta}^{\star} = 
({p_{\Delta}^{\star}}^2)^{1/2}$.
In what follows we will use these effective kinematical
quantities in all expressions; it is understood that
they reduce to the original kinematical quantities
for the free Fermi-gas calculations.

In the middle and lower panel of \figref{ps_im} 
we depict the imaginary part of the
pion self-energy in function of the pion momentum 
$|\vec{q}|$ for both mean-field calculations
at the $\Delta$ kinematics : $W_{\Delta}^{\star} 
= M_{\Delta}^{\star}$,
$|\vec{p}_{\Delta}| = 200$ MeV and pion energy 
$q^0 = [({p_{\Delta}^{\star}})^0 - E_F^{\star}]/2$.

The results in the $\sigma\omega-$model for the Pauli-corrected
decay width $\Gamma^D_{\Delta}$ and the spreading width
$\Gamma^S_{\Delta}$ at the on-shell point
$W^\star_\Delta = M_{\Delta}^{\star}(\rho)$
for both parameter sets are depicted in \figref{ds_efmn}
and \figref{dw_efmn_3d}.
It is seen that the structure of the decay width hardly changes
when effective masses are introduced; only the limiting value at 
large $\mid\vec{p}_{\Delta}\mid$ now becomes density-dependent. 
Because of the stronger reduction of the effective masses
the Pauli-blocking is more pronounced using parameter set I.

The mean-field effects result in an overall reduction of 
the spreading width as compared to the Fermi-gas calculation
(see \figref{ds_phsrc_vm}).
For the relevant nuclear densities ${\rho}/{\rho_0} \leq 1.2$,
this reduction is stronger at larger densities.
It can be shown that for vanishing $\Delta$
momentum the spreading width is roughly
proportional to the integrated pion propagator divided by the
effective $\Delta$ mass.
The reduction of the effective mass explains the global
density dependence of the spreading width in this region.
For larger densities the effect of the pion propagator
makes the spreading width saturate and eventually decrease in the
mean-field models.
The mean-field model I yields a maximal spreading width at around
the equilibrium density. In the mean-field model II, the spreading
width saturates at much larger densities.

%%%%%%%%%%%%%%%%%%%%%%%%%%%%%
\subsection{Real part of the $\Delta$ self-energy. Renormalization}
\label{Real part}

In the previous sections we have obtained the imaginary part
of the $\Delta$ self-energy due to the pion dynamics, i.e.,
the sum of the Pauli-corrected decay width and the spreading 
width. This imaginary part generates a contribution to the
real part of the $\Delta$ self-energy, which can in general
be obtained via a dispersion relation.
Based on the general structure of the $\Delta$ self-energy
in~\eqref{sigma-3/2},
one can find $Re[C_{\Delta}(p_{\Delta}^{\star})]$ and
$Re[D_{\Delta}(p_{\Delta}^{\star})]$ through a dispersion 
relation in terms of
$Im[C_{\Delta}(p_{\Delta}^{\star})]$ and
$Im[D_{\Delta}(p_{\Delta}^{\star})]$ calculated for 
arbitrary $W_{\Delta}^{\star}$.
The imaginary parts of $C_{\Delta}(p_{\Delta}^{\star})$ and
$D_{\Delta}(p_{\Delta}^{\star})$ are retrieved from the
following relations
\bgea
{\Im} C_{\Delta} (p_{\Delta}^{\star}) =
-\frac{1}{4} (\, \Gamma_{\Delta}^+
+ \Gamma_{\Delta}^-  \,)\,,\;\;\;\;\;\;\;\;
{\Im} D_{\Delta} (p_{\Delta}^{\star})  =
-\frac{1}{4 W_{\Delta}^{\star}}(\, \Gamma_{\Delta}^{+}  -
\Gamma_{\Delta}^{-}   \,)      \,,
\enea
where ${\Gamma}_{\Delta}^\pm$ is defined as
\bgea
{\Gamma}_{\Delta}^\pm  &=&
-2 Im  \, \left[
\frac{1}{4}\, Tr \, \left(\, \Sigma_{\Delta}^{\mu\nu}
\,({\cal P}^{3/2})_{\nu\mu}\,
\frac{(\pm \rlap{/} {p}_{\Delta}^{\star} + W_{\Delta}^{\star})}
{2 W_{\Delta}^{\star}}\,\right)
\right]
\nn\\
&=& -2\ Im [ C_{\Delta}(p_{\Delta}^{\star}) \pm W_{\Delta}^{\star}
\, D_{\Delta}(p_{\Delta}^{\star}) ]\,.
\label{S3/2tilde}
\enea

For instance, the imaginary parts for the one-pion loop in vacuum
($W_{\Delta}^{\star}$ reduces to $W_\Delta$) are
\bge
{\Im} C_{\Delta}(p_{\Delta}) =
- \frac{f_{\pi N\Delta}^2 k_\pi^3 M_N}
{24 \pi m_{\pi}^2 {W_{\Delta}}}\,, \;\; \;\;
\ \ {\Im} D_{\Delta}(p_{\Delta}) =
- \frac{f_{\pi N\Delta}^2 k_\pi^3 E_{k_\pi} }
{24 \pi m_{\pi}^2 W_{\Delta}^2}\,.
\label{vac-CD}
\ene
The in-medium expressions for $ImC_{\Delta}$ and $ImD_{\Delta}$
(at non-zero nucleon density) are more complicated than
Eqs.~(\ref{vac-CD}), and depend on
both $W_{\Delta}^{\star}$ and $|{\vec{p}}_{\Delta}|$.

We make the assumption that an unsubtracted dispersion 
relation holds at fixed values of $|{\vec{p}}_{\Delta}|$, 
namely
\bgea
Re [C_{\Delta}({W_{\Delta}^{\star}}^2,|\vec{p}_{\Delta}|)
&+& \rlap{/}p_{\Delta}^{\star}\, D_{\Delta}
({W_{\Delta}^{\star}}^2, |\vec{p}_{\Delta}|)] =
\nn\\
\mbox{} &&\ \frac{1}{\pi}\ \int_{W_{th}^2}^{ \infty}
\frac{[\Im C_{\Delta}({W_{\Delta}'}^2,|\vec{p}_{\Delta}|)
+ \rlap{/}p_{\Delta}^{\star}
\Im D_{\Delta}({W_{\Delta}'}^2,|\vec{p}_{\Delta}|) ] }
{{W_{\Delta}'}^2- {W_{\Delta}^{\star}}^2}
f_{\Delta}^{2}({W_{\Delta}'}^2) d{   {W_{\Delta}'}^2      }\,,
\label{dispersion-rel}
\enea
where the formfactor $f_{\Delta} ({W_{\Delta}'}^2)$  is 
introduced for convergence.
Eq.(\ref{dispersion-rel}) actually implies two separate relations 
for $C_\Delta$ and $D_\Delta$.
The threshold $W_{th}$ is $M_N +m_\pi$ in vacuum, while in medium it 
is a more complicated function of masses, Fermi momentum,
and 3-momentum $|\vec{p}_\Delta |$.

The propagator is renormalized in such a way that in vacuum
it has a pole at the physical
mass $M_\Delta=1232$ MeV with a residue equal to unity,
as if the $\Delta$ were a stable particle.
These conditions give rise to the renormalized functions 
(for arbitrary $W_\Delta$)
\bgea
Re C_{\Delta}^{R} (p_{\Delta}) &=& Re C_{\Delta}(p_{\Delta}) 
- \delta M_{\Delta}
+ M_{\Delta}(1-Z^{-1})\,, \\
Re D_{\Delta}^{R} (p_{\Delta}) &=& Re D_{\Delta}(p_{\Delta}) 
- (1-Z^{-1}) \,.
\enea
%%%%%%
The mass shift (the physical mass minus the bare mass) and  wave function
renormalization constant are given respectively by
\bgea
&& \delta M_{\Delta}= [\, Re C_{\Delta}(p_{\Delta}) + M_{\Delta}
Re D_{\Delta}(p_{\Delta})\, ]|_{W_\Delta =M_\Delta} \,,
\nn \\
&& Z = \left\{1 - Re D_{\Delta}(p_{\Delta})|_{W_\Delta =M_\Delta}
- 2M_{\Delta} \left[
Re C_{\Delta}^\prime (p_{\Delta} ) + M_{\Delta}
Re D_{\Delta}^\prime (p_{\Delta}) \,
\right]|_{W_\Delta =M_\Delta} \right\}^{-1}\,,
\label{mass-shift-renorm}
\enea
with the notation $Re C_{\Delta}^\prime (p_{\Delta})
= \partial Re C_{\Delta} (p_\Delta)/ \partial W_{\Delta}^2$
and similarly for $Re D_{\Delta}^\prime (p_{\Delta}) $.
The required properties of the renormalized propagator in vacuum
are ensured by the relations
\bgea
&& ( \Sigma_\Delta^{3/2})^R|_{\rlap{/}p_{\Delta} =M_\Delta}
=\frac{\partial }
{\partial  \rlap{/}p_{\Delta} } ( \Sigma_\Delta^{3/2})^R|_{\rlap{/}p_{\Delta}
=M_\Delta} =0 \,,
\\
&& (\Sigma_{\Delta}^{3/2})^R =
C_{\Delta}^R ( p_{\Delta} ) + D_{\Delta}^R (p_{\Delta} )
\,\rlap{/}{p_{\Delta}}  \nn \,.
\enea

Finally,  the in-medium $\Delta$ propagator, which is used in the
calculations described in sect.~\ref{Compton scatt}, reads
\bgea
G^{\mu\nu}_{\Delta} (p_{\Delta}^{\star}) &=&
\left[ {\rlap{/}p_{\Delta}^{\star}} \left(1 - Re D_{\Delta}^R
(p_{\Delta}^{\star})\right)
- \left(M_{\Delta} + Re C_{\Delta}^R(p_{\Delta}^{\star})\right) 
\right.\nn\\
\mbox{} && \left.\ \  - i \left(Im C_{\Delta}^R(p_{\Delta}^{\star}) +
Im D_{\Delta}^R(p_{\Delta}^{\star}) \rlap{/}{p_{\Delta}^{\star}} \right)
\right]^{-1}
\   ({\cal{P}}^{3/2})^{\mu\nu}\,.
\eqlab{med-mod-de-prop-K-matrix}
\enea

In the calculation of the dispersion integrals we use
a similar formfactor as in \cite{Gross}, 
\bge
\label{gross}
f_{\Delta}({W_{\Delta}'}^2) = \left(\frac{
(\Lambda^2_{\Delta} - M^2_N)^2 + (M^2_{\Delta} - M^2_N)^2}
{(\Lambda^2_{\Delta} - M^2_N)^2 + (M_N^2 - 
{\tilde{W}_{\Delta}}^2)^2} \right)^2
\ene
where $\tilde{W}_{\Delta}^2 = (W_{\Delta}' 
+ M_{\Delta} - M_{\Delta}^{\star})^2$ and
the normalization is chosen such 
that $f_{\Delta}(M_\Delta^{\star 2} ) =1$ which
is appropriate for the in-medium calculation.
The cut-off parameter is taken the same as in \cite{Gross}:
  ${\Lambda}_{\Delta} = 1.506$ GeV.

The mean-field description of the nuclear medium necessitated another 
modification of the formfactor. At large values of the 
nuclear density the decrease in the effective masses of 
N and $\Delta$ results in values close to zero for the threshold invariant mass
$W_{th}$ of the dispersion integral (\ref{dispersion-rel}), if 
$|{\vec{p}}_{\Delta}|$ is large.
Since the projection operator on the spin-3/2 sector of the
$\Delta$ appearing in the calculations of ${\Gamma}_{\Delta}^\pm$, 
leading to $Im[C_{\Delta}(p_{\Delta}^{\star})]$ and
$Im[D_{\Delta}(p_{\Delta}^{\star})]$,
has negative powers of the $\Delta$ invariant mass, 
this would cause unphysical large contributions to the dispersion
integral coming from the region of the invariant mass close to zero.
We eliminated these contributions by multiplying the formfactor in
Eq. (\ref{gross}) with 
$g(\tilde{W}_{\Delta}) = \theta(M_N - \tilde{W}_{\Delta}) 
(1 - (\tilde{W}_{\Delta} - M_N)^2 / {M_N^{\star}}^2)$.
We checked that the multiplying factor $g(\tilde{W}_{\Delta})$
hardly changes the $\Delta$ real self-energy for the
values of $|{\vec{p}}_{\Delta}|$ and $\rho$ that
enter the description of the Compton cross section 
in the next section; it is added simply to extend the
$\rho$ and $|{\vec{p}}_{\Delta}|$ range of validity of the
real part of the $\Delta$ self-energy.

\section{Compton Scattering}
\label{Compton scatt}

The amplitude for the process of Compton scattering on
a finite nucleus is calculated in the impulse approximation.
We apply the so-called factorization approximation (see~\cite{Gol64},
ch.11, sect.2) which was shown to work well in pion
photoproduction~\cite{Tia80,Era90,Dre99b} and pion
scattering~\cite{Lan73,Lan76} on nuclei, in particular for the
light nuclei in section IV, where the nuclear wave function
is well described by a harmonic oscillator model.
A large part of the effects of the Fermi-motion are
accounted for by evaluating the amplitude on a nucleon moving with
the effective momentum $p$ ($p'=p+q$) in the initial (final) state,
where $q=k -k' $ is the momentum transfer (see Appendix B for precise
definitions). The amplitude in this approximation is written as
\beq
K_A = A \, \langle \,{T}_N (\vec{p} ) \, \rangle \; F_\rho(q) \,,
\eqlab{KA}
\eeq
where $F_\rho(q)$ is the Fourier-transform of the density distribution
(formfactor). In~\eqref{KA}, 
the formfactor of the $1s$-$1p$-shell nuclei with $Z=N=A/2$ is 
constructed on the basis of the experimental charge densities
in \cite{Jager} (see table V therein), correcting for proton finite
size effects and assuming equal proton and neutron densities.
$\langle {T}_N \rangle$ is the spin averaged
single-nucleon amplitude defined as
\bea
\langle \,{T_N(\vec{p}) }\, \rangle = \frac{1}{2} \sum_{m_s = \pm 1/2}
<\,\vec{k}', \lambda'_\gamma; \vec{p}', m_s |\, T_N
|\vec{k}, \lambda_\gamma; \vec{p}, m_s \,>\,,
\eqlab{average}
\eea
where $\vec{p}$ is the effective nucleon momentum in~\eqref{solution}
and $m_s$ is the projection of the nucleon spin on the $OZ$ axis. For
isospin-saturated systems (to which we restrict our present discussion)
an isospin average is also performed.

The cross section for unpolarized photons in the Lab frame can 
now be expressed as
\bea
\frac{d \sigma}{ d \Omega'}|_{Lab}  =\frac{k'^3 E'_A}{4 \pi^2 k M_A}\;
\frac{1}{2} \sum_{\lambda_\gamma, \lambda'_\gamma}\; |K_A|^2\,,
\eqlab{cs-1}
\eea
where $E'_A = \omega + M_A$, $\omega =k-k'$ and  $\lambda_\gamma,
\lambda'_\gamma$ are the photon helicities.
It is convenient to redefine the Compton T-matrix through the
amplitude ${\cal T}_N (\vec{p})$
\bea
T_N(\vec{p}) = \Big(\, \frac{M_N }{E_N (\vec{p})}
\frac{M_N }{E_N (\vec{p}')} \frac{1}{2 k} \frac{1}{2k'} \, \Big)^{1/2}
{\cal T}_N(\vec{p})\,.
\eqlab{W_A}
\eea
The latter is normalized according
to \cite{Itz80} (App.A.3) and has simpler properties under Lorentz boosts.
Calculating the amplitude in the ${\gamma}N$ CM frame
 (marked with superscript ``$c$")
we obtain
\bea
\frac{d \sigma}{ d \Omega'}|_{Lab} =\frac{1}{16 \pi^2} \Big(\frac{k'}{k}
\Big)^2\,
\frac{E'_A}{M_A}
\frac{M_N^2}{E_N (\vec{p}) E_N (\vec{p}+\vec{q}) } A^2 F_{\rho}^2 (\vec{q})
\; \frac{1}{2} \sum_{\lambda_\gamma, \lambda'_\gamma}\;
|\langle \, {\cal T}_N(\vec{p}\,^c ) \, \rangle|^2  \,.
\eqlab{cs-2}
\eea

The photon asymmetry which can be measured with a linearly-polarized
photon beam is defined as
\bea
\Sigma_\gamma = \frac{ d\,\sigma_\perp - d\,\sigma_\parallel }{
d\,\sigma_\perp + d\,\sigma_\parallel } \,,
\eqlab{asymmetry}
\eea
where $d\,\sigma_\parallel $ ($ d\,\sigma_\perp $) is the cross section
for the photon polarization vector in the scattering plane
(perpendicular to it).

The cross section in the $\gamma\, A$ center-of-mass
frame can be related to the cross section in the Lab frame as
\bea
\frac{d \sigma}{ d \Omega' } \mid _{{}_{cm}}
= \Big(\frac{k}{k'} \Big)^2\,
\frac{M_A^2}{s_A}\; \frac{d \sigma}{ d \Omega'} \mid _{{}_{Lab}}\,,
\eqlab{cs-3}
\eea
where $s_A = M_A (M_A + 2k)$ is the total invariant energy squared, and
the center-off-mass photon momentum and the scattering angle are
\bea
k_{cm} = k M_A/ \sqrt{s_A}\,,\;\;\;\;\;\;\;\;\;\;\;\;
(\cos{\theta_\gamma})_{cm} =1- \frac{k k'}{ k^{2}_{cm} }
(1-\cos{\theta_\gamma})\,.
\eqlab{cm-kin}
\eea

The single-nucleon amplitude is decomposed into one part which
corresponds to the amplitude on the free nucleon, plus a term which
accounts for the modification of the \Del\ resonance in the medium, i.e.\
\beq
T_N = T^{free}_N + \left( K^{\Delta_d}_N - K^{\Delta_f}_N \right)
\;.
\eqlab{amplitude1}
\eeq
The first term is the $T$ matrix for Compton scattering on the free
nucleon; the term
between brackets accounts for the nuclear-medium modification of the
\Del\ resonance. To avoid double counting the vacuum contribution is
subtracted.

The T-matrix for Compton scattering off a free proton, $T^{free}_N$, is calculated in a
K-matrix model very similar to that  of ref.~\cite{Kor98}. This covariant
coupled-channels calculation of pion scattering, pion photoproduction and
Compton scattering on the nucleon, satisfies unitarity constraints
below the two-pion production threshold and is gauge invariant.
In the calculation of $T^{free}_N$ the \Del\ is treated as a genuine
spin-3/2 resonance \cite{Pas01} in order to be compatible with the present
treatment of the in-medium \Del\ resonance. The change in the structure of the
$\gamma N \Delta$ and $\pi N \Delta$ vertices
(the disappearance of the spin-1/2 off-shell couplings)
necessitated modification of parameters of the $\rho$ and
$\sigma$ exchanges in the t - channel.
A comparable fit to the data as in
ref.~\cite{Kor98} could be obtained. In~\figref{n-comp} the results for
Compton scattering are compared to data. At the pion-production
threshold, E$_{\gamma} \approx 150$MeV, the calculation overestimates
the data which might be related to ignoring in the K-matrix calculation the
real pion-loop contributions
which are responsible for the cusp structure in the $f_{EE}^{1-}$
Compton multipole~\cite{Ber93,Kon00}.

The dressed $\Delta$ contribution $K^{\Delta_d}_N$ is based on a
calculation in which only the s-type tree-level contribution is taken
into account using the medium-modified $\Delta$ propagator as defined in
sect.II (see \eqref{med-mod-de-prop-K-matrix}).
Note that in the impulse approximation
the photon is absorbed on a free nucleon and thus one has to work with the
free nucleon mass $M_N$ instead of the medium-modified one $M^{\star}_N$.
The $\Delta$ self-energy parameters will depend on the difference
between the $\Delta$ invariant mass and the nucleon mass ($\delta =W_\Delta-M_N$) and
are therefore evaluated at $W^{\star}_\Delta=M^{\star}_N + \delta$.
The subtracted $\Delta$ contribution $K^{\Delta_f}_N$ in vacuum is obtained
from a similar calculation using the free propagator instead.

In the limit of low photon energies the cross section for Compton
scattering is given by the Thomson limit, where the matrix element is
proportional to $Z^2/A$. In the present calculation only the
contribution to Compton scattering proportional to $Z$, the
total number of protons, is taken into account, thus the
contribution proportional to $NZ$ is omitted. As such the Thomson limit is violated
since $Z^2/A=Z-NZ/A$. The neglected contribution is thought to arise from
intermediate excitations to collective giant dipole resonance states and
be related to the finite extent of the nuclear system\cite{Dre84,Hut98}.
For this reason one expects this contribution to be vanishingly small at
forward angles. At backward angles, where the one-proton cross section
is suppressed by the formfactor due to the large momentum transfer, the
two-body mechanism may give a significant contribution. In our approach
the equivalent contribution would arise from two-body contributions to
the electromagnetic current arising from the nucleon-nucleon
interaction. As argued, such a contribution is of marginal importance at
forward angles but large at backward angles. In a future work this will
be included explicitly, presently we have ignored these two-body
currents.

\section{Results for coherent Compton scattering}

Cross sections have been calculated for $^4$He and  $^{12}$C at several
densities to investigate medium effects. To compare with data an average
over density ($\rho_A$), based on the  Local
Density Approximation (LDA), has been performed. The density profile
($\rho_A$) was taken consistently with the formfactor in \eqref{KA}.

In~\figref{He206} we have plotted, for various nuclear densities,
the cross section and photon asymmetry for Compton scattering
on $^4$He in mean-field model I, both at fixed lab angle
${\theta}_{\gamma} = 37^o$ and lab energy
$E_{\gamma} = 206$ MeV. The results show a strong density
dependence.
In order to obtain more insight we have plotted in
the upper panels of~\figref{par-u} the values of the
3-momentum $|\vec{p}_{\Delta}|$ and 
(kinematical) invariant mass $W_{\Delta}$ of the
${\Delta}$ as enter in the calculations presented in~\figref{He206}.
In the lower panels we show the real and imaginary part of the
$\Delta$ self-energy.
We concentrate on the dominant imaginary part,
as this seems sufficient to explain the global density dependence of
the cross section.
At fixed $E_{{\gamma}} = 206$ MeV by far the largest contribution 
to the imaginary part is due to the spreading width. 
The decay width in this energy
regime vanishes for the larger densities and is very small
for the lowest densities. The width therefore almost vanishes at 
zero density. At fixed ${\theta}_{{\gamma}} = 37^o$ and
with increasing energy the Pauli-corrected decay
width becomes more important, showing up in the $\Delta$ 
width at small density and in the global increase with energy 
starting at 300 MeV.

Much of the density dependence of the cross sections in~\figref{He206} 
can be understood from the density dependence of
the imaginary part of the $\Delta$ self-energy. 
At a photon energy of 206 MeV one is
relatively far from the peak of the $\Delta$ resonance. An increase in the
width of the resonance therefore results in an increase of the cross
section at this energy. The opposite happens when one approaches the peak
of the resonance, where the cross section decreases with density.
The data show clear evidence that this is indeed the correct mechanism,
at 206 MeV the vacuum
calculation falls below the data while the LDA result shows a
good correspondence with the data at forward angles. Near the resonance
the vacuum calculation
overestimates the data by a factor 2 while the LDA
result gives a much better prediction or even lies below. The sharp
fall-off of the cross section with angle is mostly due to the formfactor
which falls off strongly with increasing momentum transfer.

At backward angles the cross section is not reproduced, which is probably
due to the double-scattering contribution which is missing from the
present calculations. The photon asymmetry at 206 MeV shows only a minor
density dependence as compared to the error bars on the data.

In~\figref{He} we compare the $^{4}$He cross section and asymmetry
with LDA calculations for the Fermi-gas and 
mean-field calculations I and II. The Fermi-gas
calculation undershoots the data at small angles for 
$E_{\gamma} = 206$ MeV and at large energies for 
${\theta}_{\gamma} = 37^o$, 
and deviates from the asymmetry data points. 
The mean field calculations tend to improve this. 

The cross sections for $^{12}$C is shown in~\figref{C}. Because of the
larger radius of $^{12}$C the cross section falls off faster with angle
than that for $^4$He.
The drop in the cross section at energies beyond 250 MeV is partly due to an
increased width of the $\Delta$ resonance and partly due to the 
formfactor cutting the cross section at larger momentum transfers. 
This effect is also seen in the data.

\section{Summary and Conclusions}

In this paper we have presented a calculation of the cross section for 
Compton scattering on $^{4}$He and $^{12}$C in a K-matrix model
where the amplitudes are calculated in the impulse approximation. 
Fermi-motion is incorporated using the factorization approximation
scheme.
The medium effects are included by replacing the free $\Delta$ propagator
by a medium-modified $\Delta$ propagator in the s-type resonant
tree-level diagram. 
The medium-properties of the $\Delta$ resonance are investigated 
in a relativistic framework for symmetrical nuclear matter.

This involves the calculation of the $\Delta$ self-energy 
which was performed in a Fermi-gas model and two mean-field models.
The imaginary part includes the Pauli-corrected decay width 
and the spreading width incorporating ph-excitations;
the real part was calculated using dispersion integrals.
In both the Fermi-gas and mean-field calculations 
the width is increased as compared to the free
$\Delta$ width. This increase tends to be stronger for the 
Fermi-gas model than for the mean-field models.

The differential Compton scattering cross section shows 
a strong density dependence.
Within a Local Density Approximation, the density dependence
of the $\Delta$ propagator results in a much better description 
of the data, as compared to a calculation using the vacuum
$\Delta$ propagator. Mean-field models which incorporate a
reduced effective mass of the nucleon and $\Delta$ tend to improve
on the results in a Fermi-gas calculation.
Both mean-field results are quite close
indicating that the cross section is not that sensitive
to the used effective masses. 

The present one-body mechanism is unable to describe the data
at backward angles. In order to improve this, it is
imperative that multiple scattering should be incorporated 
in the model.

\section*{Acknowledgments}

O.S. thanks the Stichting voor Fundamenteel Onderzoek der Materie
(FOM) for their financial support.
A.Yu. K. thanks the Foundation for Fundamental
Research of the Netherlands (NWO) for financial support.
The work of L.V.D. and D.V.N. was supported by the Fund for Scientific 
Research - Flanders (FWO), the Institute for the Promotion of
Innovation by Science and Technology in Flanders (IWT)
and the Research Board of Ghent University.

\appendix

\section{    Analytical expressions for the pion self-energy}
The real and imaginary part of the pion self-energy can be expressed as
\bgea
Re {\Pi}^0_{ph} (q) &=& 8 \left(\frac{f_{{\pi}NN}}{m_{\pi}}
\right)^2 \left[|\vec{q}|^2 C_0 (|\vec{q}|) - C_2 (|\vec{q}|)
+ 2 M_N^2 q^2 Re[L_0 (q)]\right]\\
Im {\Pi}^0_{ph} (q) &=& 8 \left(\frac{f_{{\pi}NN}}{m_{\pi}}
\right)^2 \left[2 M_N^2 q^2 Im[L_0 (q)] \right]
\enea
where the expressions for the functions $C_0$ and $C_2$ read as
\bgea
C_0(|\vec{q}|) &=& \int \frac{d\vec{k}}{(2 {\pi})^3}
\frac{\theta(|\vec{k}| -
k_F) \theta(k_F - |\vec{k} + \vec{q}|)}{4 E_{\vec{k}} E_{\vec{k} +
\vec{q}}} \left[E_{\vec{k}} - E_{\vec{k} + \vec{q}} \right] \\
&=& \left[\frac{1}{96 {{\pi}^2} |\vec{q}|} \right]
\Biggl[
\left[3(|\vec{q}|^2 - k_F^2 - |\vec{q}||\vec{k}|)
 - 2 M_N^2 + |\vec{k}|^2 \right]  E_{\vec{k}}
+ \left[|\vec{q}|^2 - 2(M_N^2 + |\vec{k}|^2)
\right. \Biggr.\nn\\
\mbox{} && \left.\Biggl.\left. + |\vec{q}||\vec{k}| \right]
E_{|\vec{k}| - |\vec{q}|}
+ 3 E_F |\vec{k}|^2 + 3 {|\vec{q}| M_N^2}
ln \left(\frac{(E_{\vec{k}} + |\vec{k}|)}
{(E_{|\vec{k}| - |\vec{q}|} + |\vec{k}| - |\vec{q}|)}
\right) \Biggr]
\right|_{|\vec{k}| = k_1}^{|\vec{k}| = k_2}
\enea
\bgea
C_2(|\vec{q}|) &=& \int \frac{d\vec{k}}{(2 {\pi})^3}
\frac{\theta(|\vec{k}| -
k_F) \theta(k_F - |\vec{k} + \vec{q}|)}{4 E_{\vec{k}} E_{\vec{k} +
\vec{q}}} \left[E_{\vec{k}} - E_{\vec{k} + \vec{q}} \right]
\left[E_{\vec{k}} + E_{\vec{k} + \vec{q}} \right]^2 \\
&=& \left[\frac{1}{16 {{\pi}^2} |\vec{q}|} \right]
\Biggl[
(2M_N^2 - k_F^2) \frac{E_F}{6} |\vec{k}|^2
+ \frac{E_F}{4} |\vec{k}|^4
+ \frac{1}{60} \left[15|\vec{q}|^4 - 40 M_N^2 |\vec{q}|^2
- 8 M_N^4  \right. \Biggr. \nn\\
\mbox{} && \left. -\ 15 k_F^4 - 20 M_N^2 k_F^2 - 30 |\vec{q}|^3|\vec{k}|
+ 10 k_F^2|\vec{k}|^2 + 20 |\vec{q}|^2|\vec{k}|^2
+ 4 M_N^2 |\vec{k}|^2 - 3 |\vec{k}|^4
\right] E_{\vec{k}} \nn\\
\mbox{} && -\ \frac{1}{30}\left[4|\vec{k}|^4 - |\vec{q}|^4
+ 4 M_N^4 + 4|\vec{q}||\vec{k}|^3 - 6 |\vec{q}|^2|\vec{k}|^2
- |\vec{q}|^3|\vec{k}| + 8 M_N^2|\vec{k}|^2
\right. \nn\\
\mbox{} && \left. \Biggl.\left. +\ 18 M_N^2 |\vec{q}|^2
+ 4 M_N^2|\vec{q}||\vec{k}| \right] E_{|\vec{k}| - |\vec{q}|}
+ \frac{|\vec{q}|^3 M_N^2}{2}
\ ln \left(\frac{(E_{|\vec{k}|} + |\vec{k}|)}
{E_{|\vec{k}| - |\vec{q}|} + |\vec{k}| - |\vec{q}|}\right)
\Biggr] \right|_{|\vec{k}| = k_1}^{|\vec{k}| = k_2}
\enea
with the integration boundaries
\bge
k_1 = \mbox{max}(k_F,|\vec{q}| - k_F)\ \ \ ,\ \ \ \  k_2 = k_F + |\vec{q}|.
\ene
The relativistic equivalent of the Lindhard function is
\bge
L_0(q) = \int \frac{d\vec{k}}{(2 {\pi})^3} \frac{\theta(|\vec{k}| -
k_F) \theta(k_F - |\vec{k} + \vec{q}|)}{4 E_{\vec{k}} E_{\vec{k} +
\vec{q}}} \left[\frac{1}{E_{\vec{k}} - E_{\vec{k} + \vec{q}}
- q_0 - i\epsilon} - \frac{1}{E_{\vec{k}} - E_{\vec{k} + \vec{q}}
+ q_0 - i\epsilon} \right].
\ene
Based on the work in \cite{Kuras1,Liu2} the real and
imaginary part of $L_0$ can be written as\footnote{In the original
article \cite{Liu2} some typing errors occur in these formulas.}
\bgea
Re[L_0(q)] &=& \frac{1}{8(2{\pi})^2 |\vec{q}|} \Biggl[2 E_F I_{+}^{(-)}
(q_0) + q_0 I_{-}^{(-)}(q_0) + 2 \left(E_{+}^{(-)} - E_{-}^{(-)}
\right) \Biggr. \nn\\
\mbox{} && -\ \sqrt{{\alpha}} \left[ \theta(q^2 - 4 M_N^2)
+ \theta(- q^2) \right]
I_{-}^{(+)}(\sqrt{{\alpha}}) +\ |\vec{q}| I_{-}^{(+)}(|\vec{q}|) \nn\\
\mbox{} && \left. +\ 2 \sqrt{-{\alpha}}\ 
\theta (q^2)\ \theta (4M_N^2 -q^2)
\left[ arctg \left(\frac{E_{+}^{(+)}}{\sqrt{-{\alpha}}}\right)
- arctg \left(\frac{E_{-}^{(+)}}{\sqrt{-{\alpha}}}\right) \right]
\right]\\
Im[L_0(q)] &=& \frac{{\pi}}{8(2{\pi})^2 |\vec{q}|}
\left[[(2 E_F + q_0)
- f(q_0)] 
\ \theta(q_0 -\mbox{max}(0,E_{-}^{(-)}) )\ \theta(E_+^{(-)} - q_0)
\right. \nn\\
\mbox{} &&
\left. -\  \theta(2 k_F - |\vec{q}|)\ [(2 E_F - q_0) - f(q_0)]
\ \theta(q_0)\ \theta(- E_{-}^{(-)} - q_0)
\right]
\enea
where
\bgea
I_{+}^{(\pm)} (x) &=& ln \left[\frac{(E_{+}^{(\pm)} + x)
(E_{+}^{(\pm)} - x)}{(E_{-}^{(\pm)} + x)(E_{-}^{(\pm)} - x)} \right];\
I_{-}^{(\pm)} (x) = ln \left[\frac{(E_{-}^{(\pm)} + x)
(E_{+}^{(\pm)} - x)}{(E_{+}^{(\pm)} + x)(E_{-}^{(\pm)} - x)} \right]\\
E_{+}^{(\pm)} &=& E_{k_F + |\vec{q}|} \pm E_F\ \ ;\ \ \
E_{-}^{(\pm)} = E_{k_F - |\vec{q}|} \pm E_F\\
f(x) &=& |\vec{q}| \sqrt{1 + \frac{4 M_N^2}{|\vec{q}|^2 - x^2}}\ \ ;\ \ \
\alpha = f^2(q_0)\\
E_{|\vec{k}| \pm |\vec{q}|} &=& \sqrt{|\vec{k}|^2 \pm
2|\vec{k}||\vec{q}| + |\vec{q}|^2 + M_N^2} .
\enea
All analytic expressions have also been checked numerically.

\section{Kinematics}

%\subsection{Laboratory frame }

We consider kinematics in the laboratory frame for the $\gamma\; A$
scattering, where the initial nucleus is at rest,
\bea
\omega+M_A = \sqrt{M_A^2 +\vec{q}^2}
\label{nucleus}
\eea
and $(\omega,\vec{q})=(k-k', \vec{k}-\vec{k}')$ is the 4-momentum
transferred to the nucleus $A$. The energy of the final photon is
\bea
k' = \frac{k}{ 1+k/M_A \,(1-\cos{\theta_\gamma}) }
\eea
where $\theta_\gamma$ is the photon scattering angle.

The nucleon ``effective'' momentum $\vec{p}$ can be found by assuming energy-momentum
conservation on the constituent nucleon
\bea
\omega+\sqrt{M_N^2+\vec{p}^2} =\sqrt{M_N^2 +(\vec{p}+\vec{q})^2}
\label{nucleon}
\eea
with the same 4-momentum transfer.

In the non-relativistic approximation, where $|\vec{q}| \ll m$, one
obtains from Eqs.~(\ref{nucleus}) and (\ref{nucleon})
\bea
2 \vec{p}\vec{q}= {\vec{q}}^{\,2} (\frac{M_N}{M_A}-1)\,.
\eqlab{nonrel}
\eea
A possible solution of \eqref{nonrel} is given by
\bea
\vec{p}= - \alpha \vec{q}\,
\eqlab{solution}
\eea
with
\bea
\alpha= \frac{1}{2} (1-\frac{M_N}{M_A}) \approx
\frac{1}{2} (1-\frac{1}{A}     )\,.
\eqlab{nr-alpha}
\eea
The component of the momentum $\vec{p}$ perpendicular to $\vec{q}$
is not determined from \eqref{nonrel} and may be conveniently
chosen equal to zero.

In the relativistic case the solution of Eq.~(\ref{nucleon}) 
is more complicated.
If we seek
for an effective momentum in the form of \eqref{solution} then
we obtain two solutions
\bea
\alpha_{\pm} =\frac{1}{2}\Big( 1 \pm \frac{\omega}{|\vec{q}|}
\sqrt{1+2\frac{M_N^2}{\omega M_A} } \;               \Big) \,.
\eqlab{rel-alpha}
\eea
It is easy to check that for the non-relativistic kinematics
$\alpha_-$ in \eqref{rel-alpha} reduces to \eqref{nr-alpha}.
Forward scattering ($\theta_\gamma =0$) is a special case for which
$\omega = \vec{q} =0$. In this case the exact solution is given by
\eqref{nr-alpha}.

To make a transformation of the single-nucleon T-matrix from the
system where the nucleon moves with a momentum $\vec{p}$
to the $\gamma\; N$ center-of-mass system  we consider the invariant
Mandelstam variables
\bea
s_N =(k+p)^2 = M_N^2+2k E_N (\vec{p})-2\vec{k}\vec{p}\,, \nonumber\\
t_N =(k-k')^2 = -2k k' (1-\cos{\theta_\gamma})\,.
\eea
The corresponding center-of-mass 3-momentum and photon scattering angle 
can be obtained from
\bea
|\vec{p}^c|=|\vec{k}^c|=(s_N - M_N^2)/2\sqrt{s_N}\,,
\;\;\;\;\;\;\;\;\;\;\;\;
\cos{\theta_N^c} =1+t_N / 2 (\vec{k}^c)^2\,.
\eea

%%%%%%%%%%%%%%%%%%%%%%%%%%%%%%%%%%%%%%%%%
%%%% Figures : %%%%%%%%%%%%%%%%%%%%%%%%%%
%%%%%%%%%%%%%%%%%%%%%%%%%%%%%%%%%%%%%%%%%

%%Fig1:%%

\begin{figure}[hbt]
\begin{center}
\leavevmode
\epsfysize=15cm
%%{\epsfbox{ds_dec_fg_2.ps}}
{\epsfbox{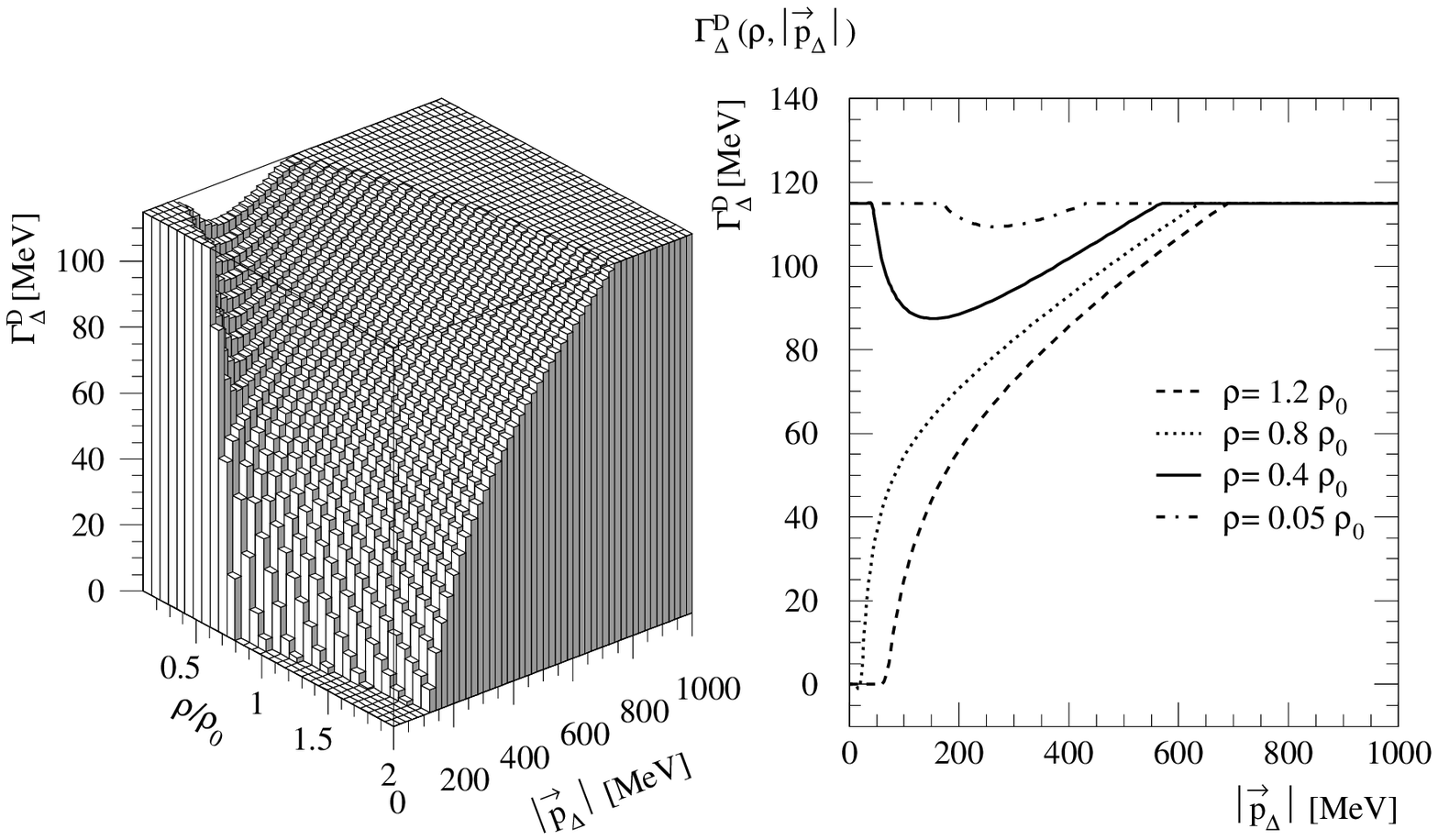}}
\caption[dns]{(a) In the left panel the Pauli-corrected $\Delta$ decay
width is depicted as a function of the density $\rho$ 
(in units of the equilibrium density ${\rho}_0$ [$k_F^0 = 1.333 fm^{-1}$])
and the $\Delta$ 3-momentum $|{\vec{p}}_{\Delta}|$ at $W_\Delta
= 1232$ MeV calculated in the Fermi-gas model. (b) The right panel
shows the results for this calculation in function of 
the $\Delta$ 3-momentum $|{\vec{p}}_{\Delta}|$
for the 4 different densities stated in the figure.\\
}
\figlab{ds_pc_vm}
\end{center}
\end{figure}

%%Fig2:%%

\begin{figure}[hbt]
\begin{center}
\leavevmode
\epsfysize=15cm
%%{\epsfbox{ps_im.ps}}
{\epsfbox{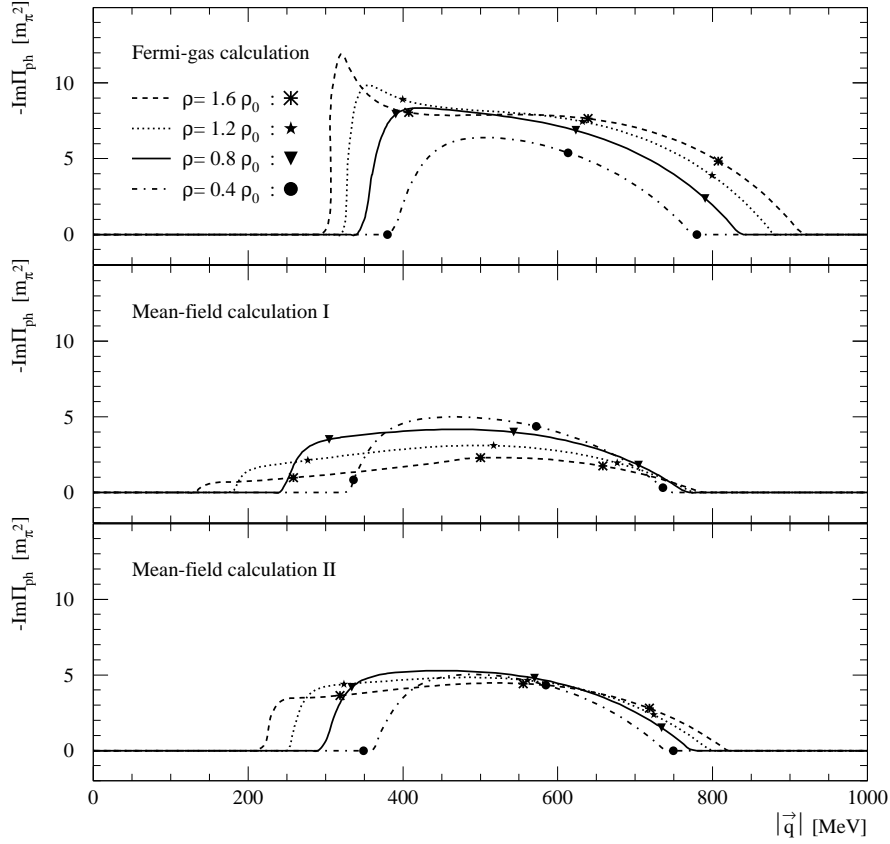}}
\caption[dns]{The imaginary part of the pion self-energy
in function of the pion 3-momentum $|{\vec{q}}|$ at 
$q^0 = [(p_{\Delta})^0 - E_F] / 2$
for the ${\Delta}$ kinematics $W_{\Delta} = 1232$ MeV
and $|\vec{p}_{\Delta}| = 200$ MeV
within the Fermi-gas model and at 
$q^0 = [({p_{\Delta}^{\star}})^0 - E_F^{\star}] / 2$
for the ${\Delta}$ kinematics $W_{\Delta}^{\star} = M_{\Delta}^{\star}$
and $|\vec{p}_{\Delta}| = 200$ MeV for the mean-field models I and II.
The markers on the figures indicate the $|\vec{q}|$-values for 
$cos{\theta}$ equal to -1 (first marker), 0 (second) and 1 (third) for 
the specific kinematics stated above
which appear in the spreading width calculation,
see \eqref{spread}.\\
}
\figlab{ps_im}
\end{center}
\end{figure}

%%Fig3:%%

\begin{figure}[hbt]
\begin{center}
\leavevmode
\epsfysize=15cm
%%{\epsfbox{ds_spr_fg_2.ps}}
{\epsfbox{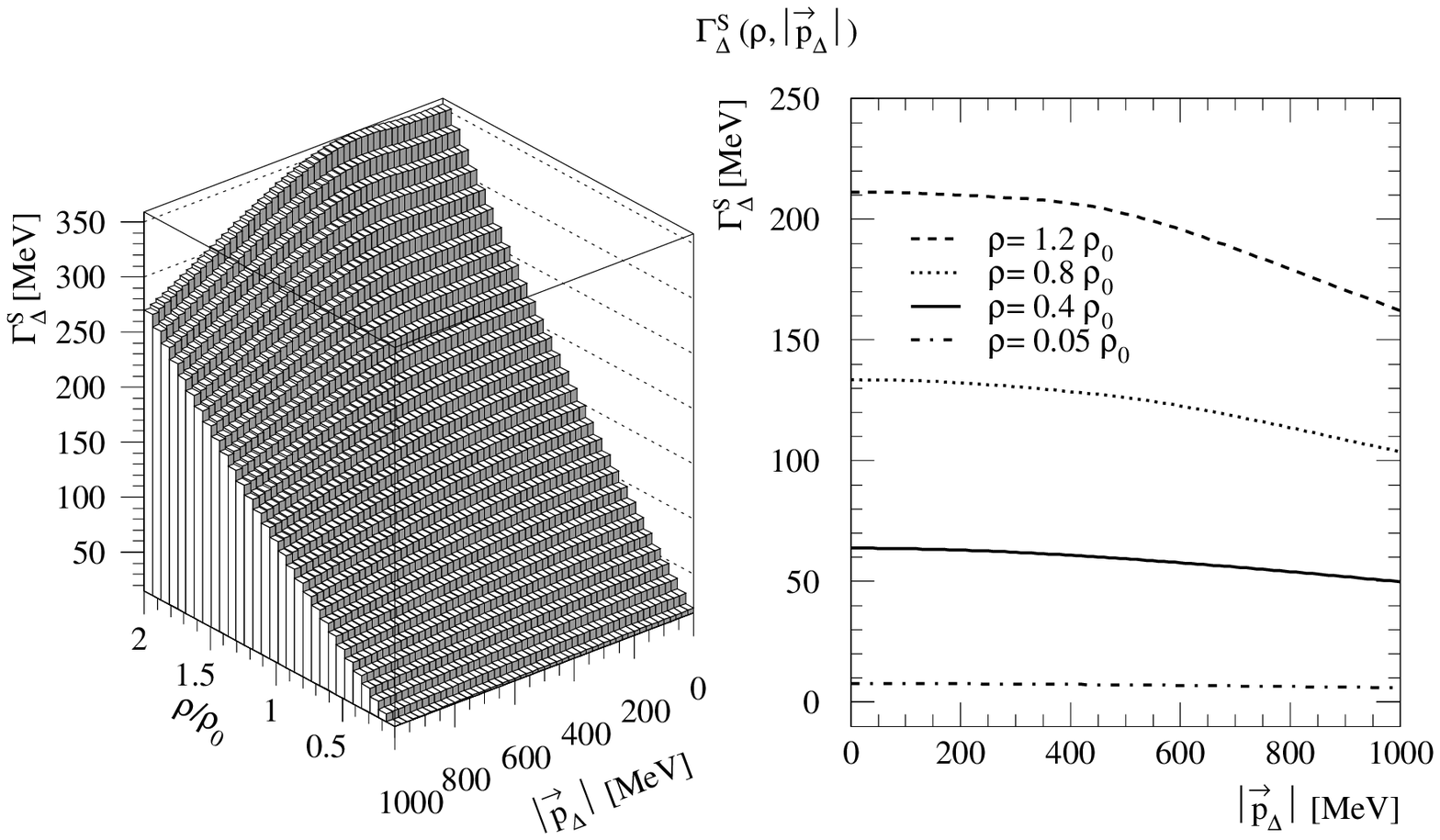}}
\caption[dns]{(a) In the left panel the spreading width is depicted as 
a function of the density $\rho$ 
(in units of the equilibrium density ${\rho}_0$ [$k_F^0 = 1.333 fm^{-1}$])
and the $\Delta$ 3-momentum $|{\vec{p}}_{\Delta}|$ at $W_\Delta
= 1232$ MeV calculated in the Fermi-gas model. (b) The right panel
shows the results for this calculation in function of 
the $\Delta$ 3-momentum $|{\vec{p}}_{\Delta}|$
for the 4 different densities stated in the figure.\\
}
\figlab{ds_phsrc_vm}
\end{center}
\end{figure}

%%Fig4:%%

\begin{figure}[hbt]
\begin{center}
\leavevmode
\epsfysize=15cm
%%{\epsfbox{rho_efmn_3_4b.ps}}
{\epsfbox{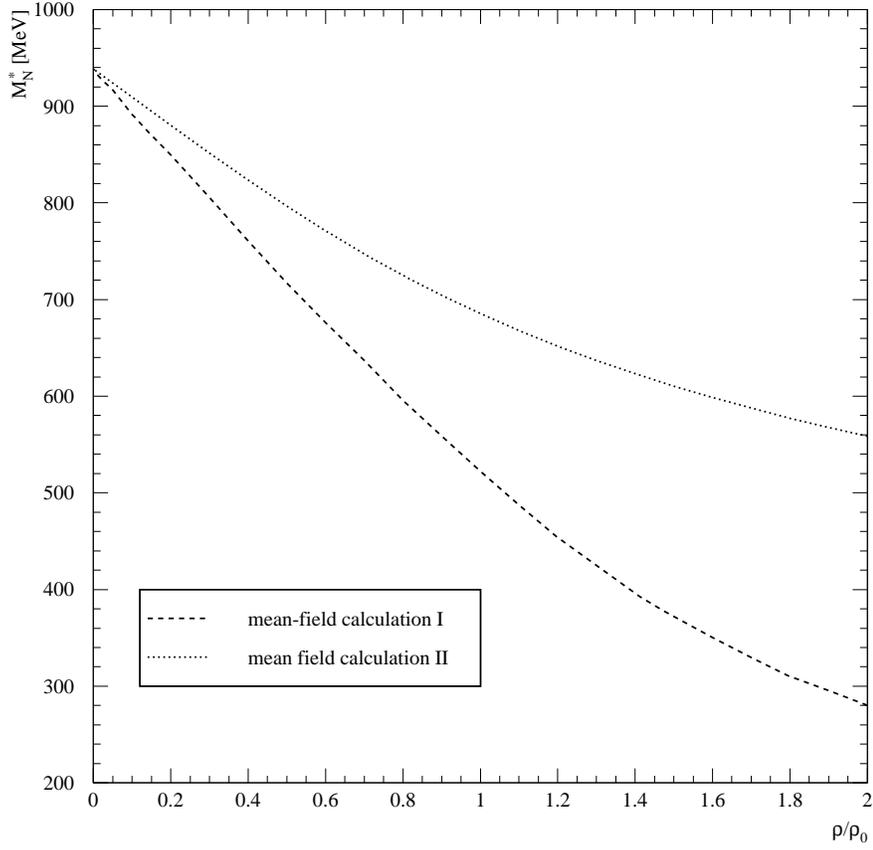}}
\caption[dns]{The effective nucleon mass $M_N^{\star}$ as a function
of the density $\rho$ 
(in units of the equilibrium density ${\rho}_0$ (value depending 
on the model)) for the mean-field calculations I and II 
as explained in section \ref{Effective mass}.
}
\figlab{kf_efmn_1}
\end{center}
\end{figure}

%%Fig5:%%

\begin{figure}[hbt]
\begin{center}
\leavevmode
\epsfysize=15cm
%%{\epsfbox{ds_dec_spr_3_4b.ps}}
{\epsfbox{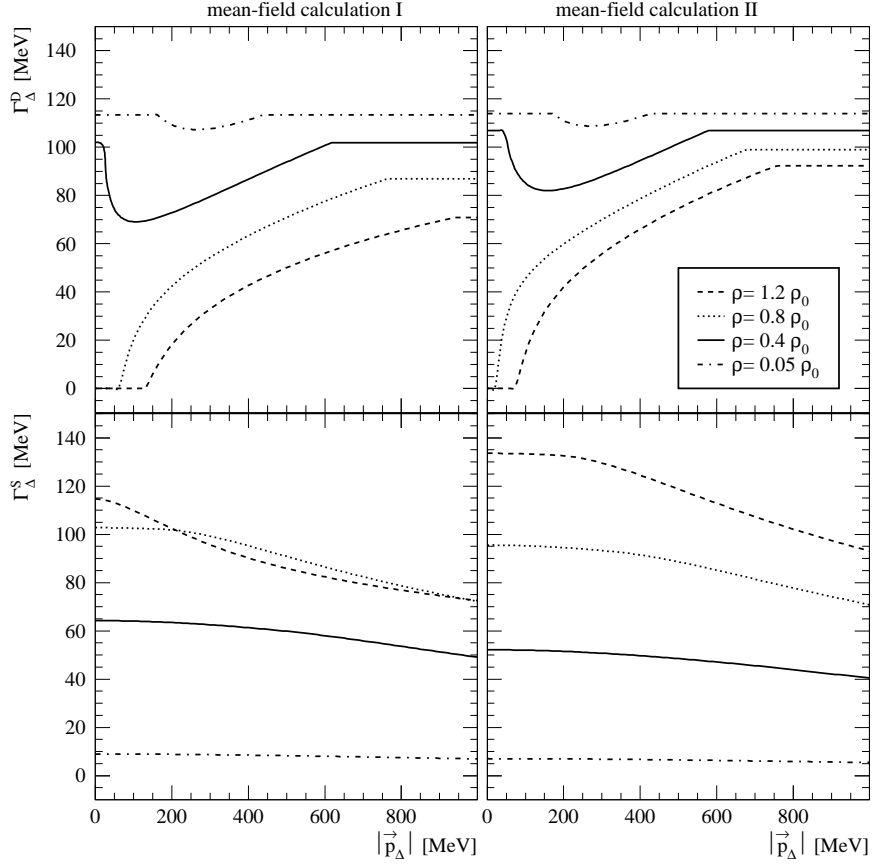}}
\caption[dns]{The Pauli-corrected decay width $\Gamma_{\Delta}^{D}$ and 
the spreading width $\Gamma_{\Delta}^{S}$ as a function of the
$\Delta$ 3-momentum $|{\vec{p}}_{\Delta}|$ calculated in the $\sigma
\omega-$model  at $W^\star_\Delta =
M_{\Delta}^{\star}(\rho)$ for 4 different densities and for
the two mean-field calculations as explained in the text.
}
\figlab{ds_efmn}
\end{center}
\end{figure}

%%Fig6:%%

\begin{figure}[hbt]
\begin{center}
\leavevmode
\epsfysize=15cm
%%{\epsfbox{dw_efm_3d_2-fl_2.ps}}
{\epsfbox{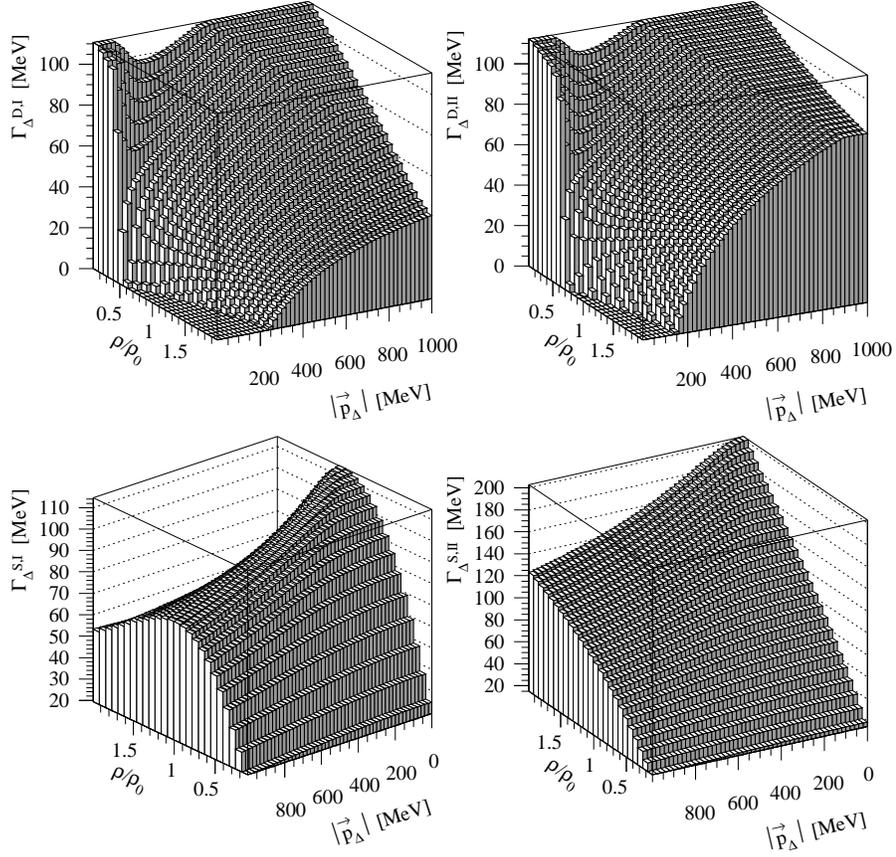}}
\caption[dns]{The Pauli-corrected $\Delta$ decay width $\Gamma^D_{\Delta}$
and the spreading width $\Gamma^S_{\Delta}$
in function of the density $\rho$
(in units of the equilibrium density ${\rho}_0$ (value depending 
on the model)) 
and the $\Delta$ 3-momentum $|{\vec{p}}_{\Delta}|$ calculated in the
mean-field model I and II (marked by index I and II respectively)
at $W^\star_\Delta = M_{\Delta}^{\star}(\rho)$.
}
\figlab{dw_efmn_3d}
\end{center}
\end{figure}

%%Fig7:%%

\begin{figure}[hbt]
\begin{center}
\leavevmode
\epsfysize=12cm
%%{\epsfbox{n-comp.PS}}
{\epsfbox{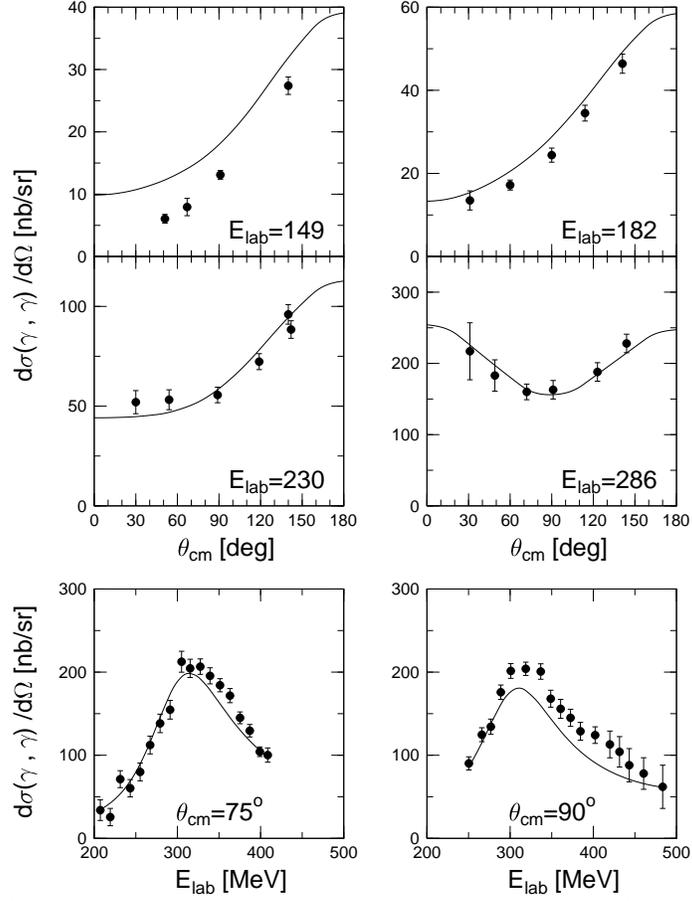}}
\caption[dns]{The calculated cross section for Compton scattering off
the proton as a function of angle at fixed photon
energy, and as a function of photon energy at fixed angle.
Data are taken from~\cite{Hal93}.
}
\figlab{n-comp}
\end{center}
\end{figure}

%%Fig8:%%

\begin{figure}[hbt]
\begin{center}
\leavevmode
\epsfysize=15cm
%%\epsfbox{HE206W4.PS}}
{\epsfbox{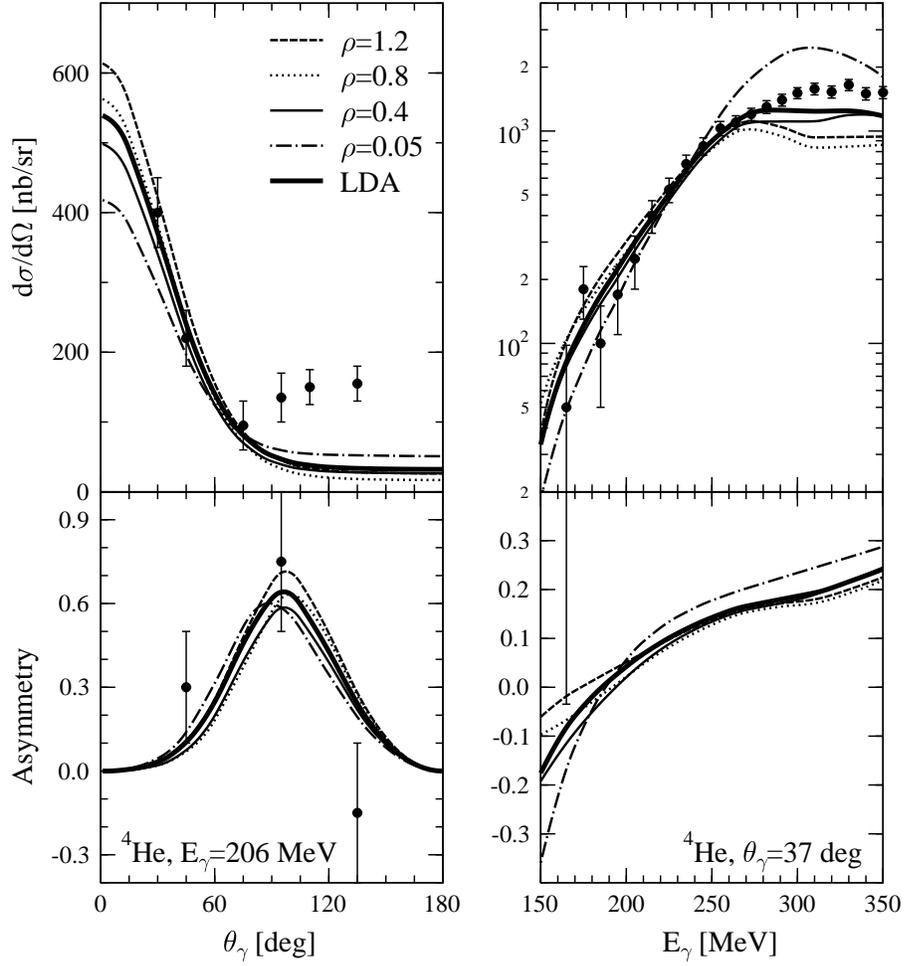}}
\caption[f-he]{Differential cross section
and photon asymmetry for Compton scattering off $^4$He at
an energy of 206 MeV as a function of angle and at an angle of 37$^o$ as a
function of energy. The curves represent the results of the Local Density
Approximation calculation (LDA) and those for different densities 
(in units of saturation density) for the mean-field calculation I.
Data are taken from~\cite{Kra98,Mor95}. }
\figlab{He206}
\end{center}
\end{figure}

%%Fig9:%%

\begin{figure}[hbt]
\begin{center}
\leavevmode
\epsfysize=15cm
%%{\epsfbox{PAR-W4.PS}}
{\epsfbox{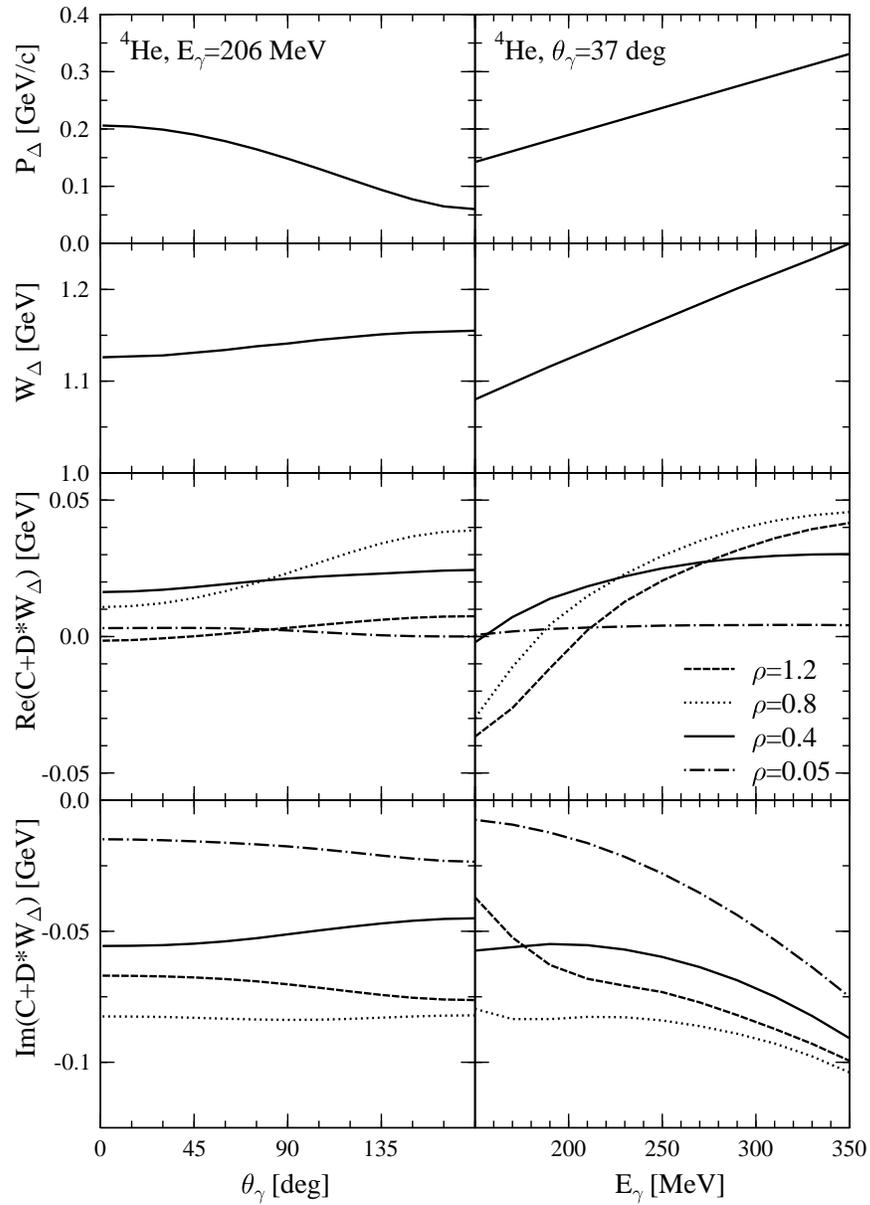}}
\caption{Values of the parameters which define the self-energy of the
$\Delta$ resonance evaluated at the $\Delta$ invariant mass 
($W_\Delta$) and three momentum ($|\vec{p}_{\Delta}|$)
appropriate for Compton scattering off $^4$He as shown in \figref{He206}.}
\figlab{par-u}
\end{center}
\end{figure}

%%Fig10:%%

\begin{figure}[hbt]
\begin{center}
\leavevmode
\epsfysize=15cm
%%{\epsfbox{HE206F1.PS}}
{\epsfbox{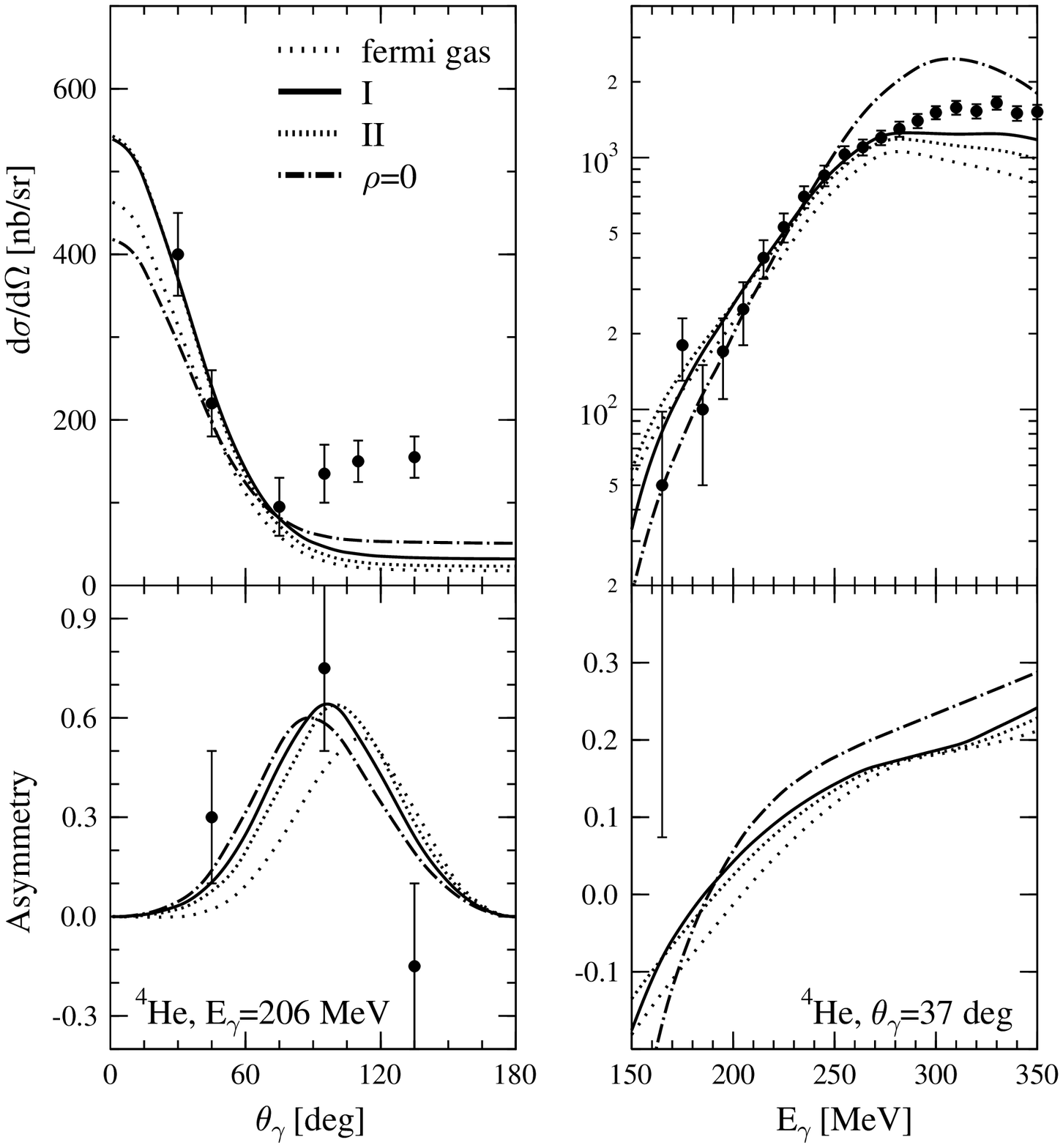}}
\caption[f-c]{LDA calculation of coherent Compton scattering off
$^{4}$He in the Fermi-gas model and the mean-field models I and II.
The data are from ref.~\cite{Kra98,Mor95}. }
\figlab{He}
\end{center}
\end{figure}

%%Fig11:%%

\begin{figure}[hbt]
\begin{center}
\leavevmode
\epsfysize=15cm
%%{\epsfbox{CF1.PS}}
{\epsfbox{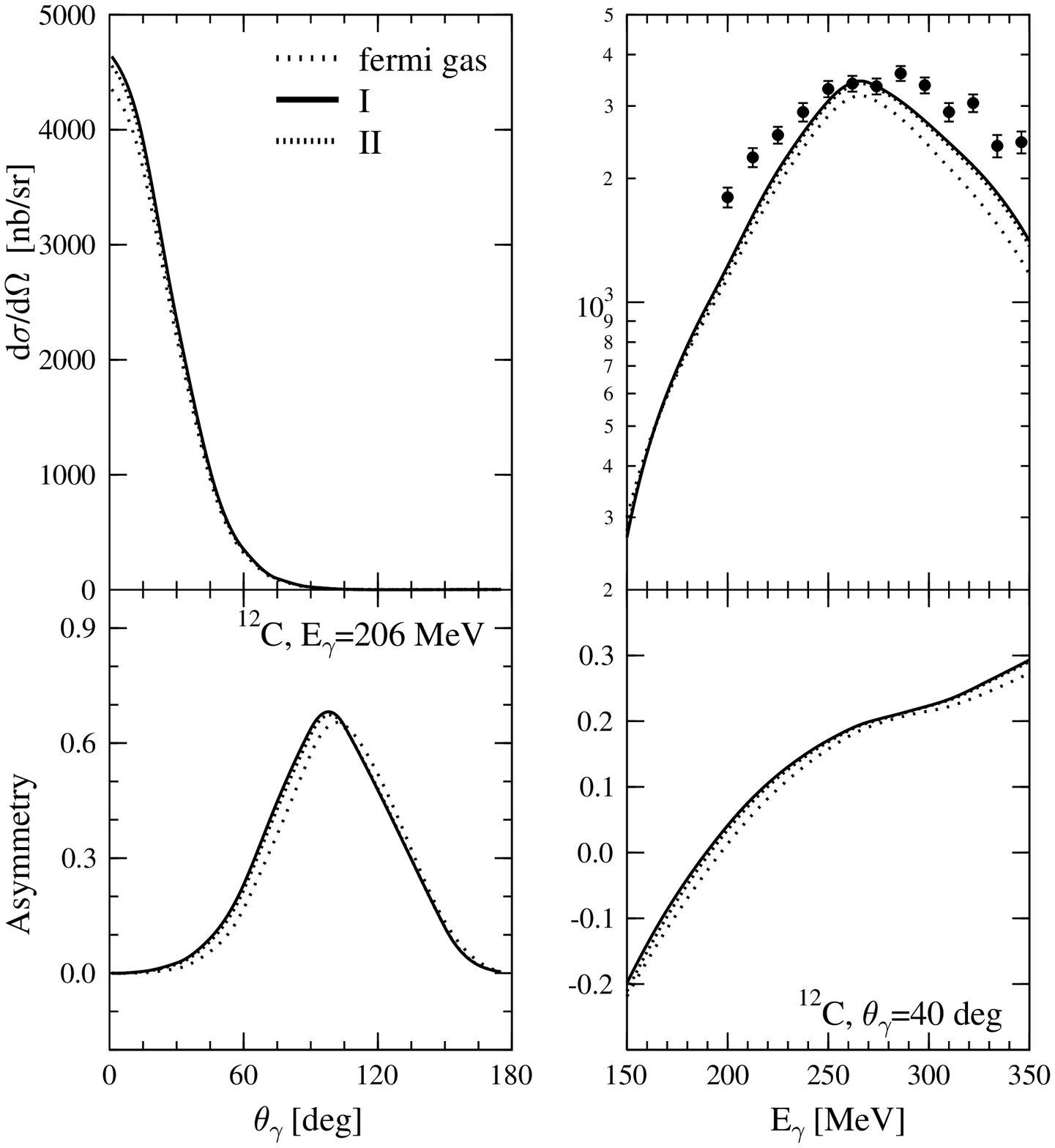}}
\caption[f-c]{LDA calculation of coherent Compton scattering off
$^{12}$C in the Fermi-gas model and the mean-field models I and II.
The data are from ref.~\cite{Wis94}. }
\figlab{C}
\end{center}
\end{figure}

\end{document}